# Approximating the Gaussian Multiple Description Rate Region Under Symmetric Distortion Constraints

Chao Tian, Soheil Mohajer, and Suhas N. Diggavi

**Abstract**

We consider multiple description coding for the Gaussian source with $K$ descriptions under the symmetric mean squared error distortion constraints, and provide an approximate characterization of the rate region. We show that the rate region can be sandwiched between two polytopes, between which the gap can be upper bounded by constants dependent on the number of descriptions, but independent of the exact distortion constraints. Underlying this result is an exact characterization of the lossless multi-level diversity source coding problem: a lossless counterpart of the MD problem. This connection provides a polytopic template for the inner and outer bounds to the rate region. In order to establish the outer bound, we generalize Ozarow's technique to introduce a strategic expansion of the original probability space by more than one random variables. For the symmetric rate case with any number of descriptions, we show that the gap between the upper bound and the lower bound for the individual description rate is no larger than 0.92 bit. The results developed in this work also suggest the "separation" approach of combining successive refinement quantization and lossless multi-level diversity coding is a competitive one, since it is only a constant away from the optimum. The results are further extended to general sources under the mean squared error distortion measure, where a similar but looser bound on the gap holds.

## 1 Introduction

In the multiple description (MD) problem, a source is encoded into several descriptions such that any one of them can be used to reconstruct the source with certain quality, and more descriptions can improve the reconstruction. The problem is well motivated by source transmission over unreliable network and distributed storage systems, since there exists uncertainty as to which transmissions are received successfully (or which servers are accessible) by the end user.

In the early works on this problem, for example [1, 2], only two descriptions are considered. Even in this setting, the quadratic Gaussian problem is the only completely solved case [2], for which the achievable region in [1] is tight. Through a counter-example, Zhang and Berger showed that this achievable region is however not tight in general [3], and a complete characterization of the rate-distortion (R-D) region has not been found to this date. See [4] (and the references therein) for a review of works related to this problem in the information theory literature.

Recent research attention has shifted to the general $K$-description problem, partly motivated by the availability of multiple transmission paths in modern communication networks. In [5][6], an achievable individual description rate was provided for symmetric multiple descriptions, where each description has the same rate, and the distortion constraint depends only on the number of descriptions available. This achievable region is based on joint binning of the codebooks for each description, which has a similar flavor as the method often used in distributed source coding problems. Another achievable region was given in [7] using more conventional conditional codebooks. Wang and Viswanath [8, 9] generalized the Gaussian MD problem to vector Gaussian source with many descriptions, and tight sum rate lower bound was established for certain cases with only *two levels* of distortion constraints (see also the outer bound result in [7]).

In this work, we consider general multiple description coding with $K$ descriptions under symmetric distortion constraints. The distortion constraints are symmetric in the sense that with any $k \leq K$ descriptions, the reconstruction has to satisfy the distortion $D_k$, regardless of which specific combination of $k$ descriptions is used. Though the distortion constraints are symmetric, the rates of the descriptions are not necessarily the same in this setting, thus generalizing the case treated in [5][6]. Nevertheless the completely symmetric case as considered in [5][6], i.e., with both symmetric rate and distortion constraints, is indeed an interesting special case, and will be treated with particular care. Our main focus is on the Gaussian source under the mean squared error (MSE) distortion constraint, however we also show that the results can be extended to more general sources under the same distortion measure.

Though completely characterizing the rate-distortion region of the Gaussian multiple description problem is difficult if not impossible, we provide an approximate characterization. Underlying this approximation is the lossless symmetric multi-level diversity (MLD) coding problem previously studied in [15, 16]; see Fig. 1. The MLD coding problem can be interpretted as a lossless version of the MD problem, and thus one of our main insights is to use the MLD rate region as a polytopic template for inner and outer bounding the MD rate-distortion region. We show that the MD rate-distortion region can be sandwiched between two polytopes, between which the gap can be upper bounded by constants dependent on the number of descriptions, but independent of the exact distortion constraints. The MD coding system is illustrated in Fig. 1 for $K = 3$ together with the MLD coding system.

One of the main contributions of this work is a novel lower bound to the sum rate for the Gaussian source, under $K$ levels of symmetric distortion constraints. This generalizes previous results in [2, 8, 9], where only two levels of distortion constraints are enforced in the system. Though the lower bound given here may not be tight, it is the first provably good bound with more than two levels of distortion constraints enforced, to the best of our knowledge. We derive this lower bound by generalizing Ozarow's technique in treating the Gaussian two-description problem. More specifically, we expand the probability space of the original problem by more than one auxiliary random variables, and impose certain Markov structure on these random variables. Ozarow's technique has been applied to various problems besides the MD problem [2, 7–9], for example, the results on multi-terminal source coding by Wagner and Anantharam [10], and the joint source channel coding problem with bandwidth expansion by Reznic *et al.* [11]. However, in all these previous works the probability space is expanded by only one additional auxiliary random variable (in [8, 9] it is one additional auxiliary random vector since vector source was being considered). Recently a similar technique has also been applied to the Gaussian interference channel problem [12], and interestingly the results there indeed require expanding the probability space by more than one random variables. The MD sum rate lower bound given in our work can be optimized over $K - 1$ variables to provide the tightest bound. However an explicit solution for this optimization problem appears difficult, thus instead we choose a specific set of values to provide a suboptimal lower bound, which nevertheless still offers insight on the problem and allows us to give an approximate characterization of the MD rate region.

For the inner bounds, we analyze two achievability schemes: the first is a very simple scheme based on successive refinement coding [13, 14] coupled with multi-level diversity coding (SR-MLD) [15–18]; the second is a generalization of the multilayer coding scheme proposed by Puri, Pradhan and Ramchandran [5][6], which we will refer to as the PPR multilayer scheme. In the special case of symmetric rate, the first scheme reduces to the well-known unequal loss protection method [20], and we thus also refer to it as the SR-ULP scheme. The SR-MLD (or SR-ULP) scheme is in fact a separation-based scheme where the quantization step and lossless source coding step are performed separately. As illustrated in Fig. 3, the output of a successive refinement code is cascaded with the lossless multi-level diversity coding scheme.

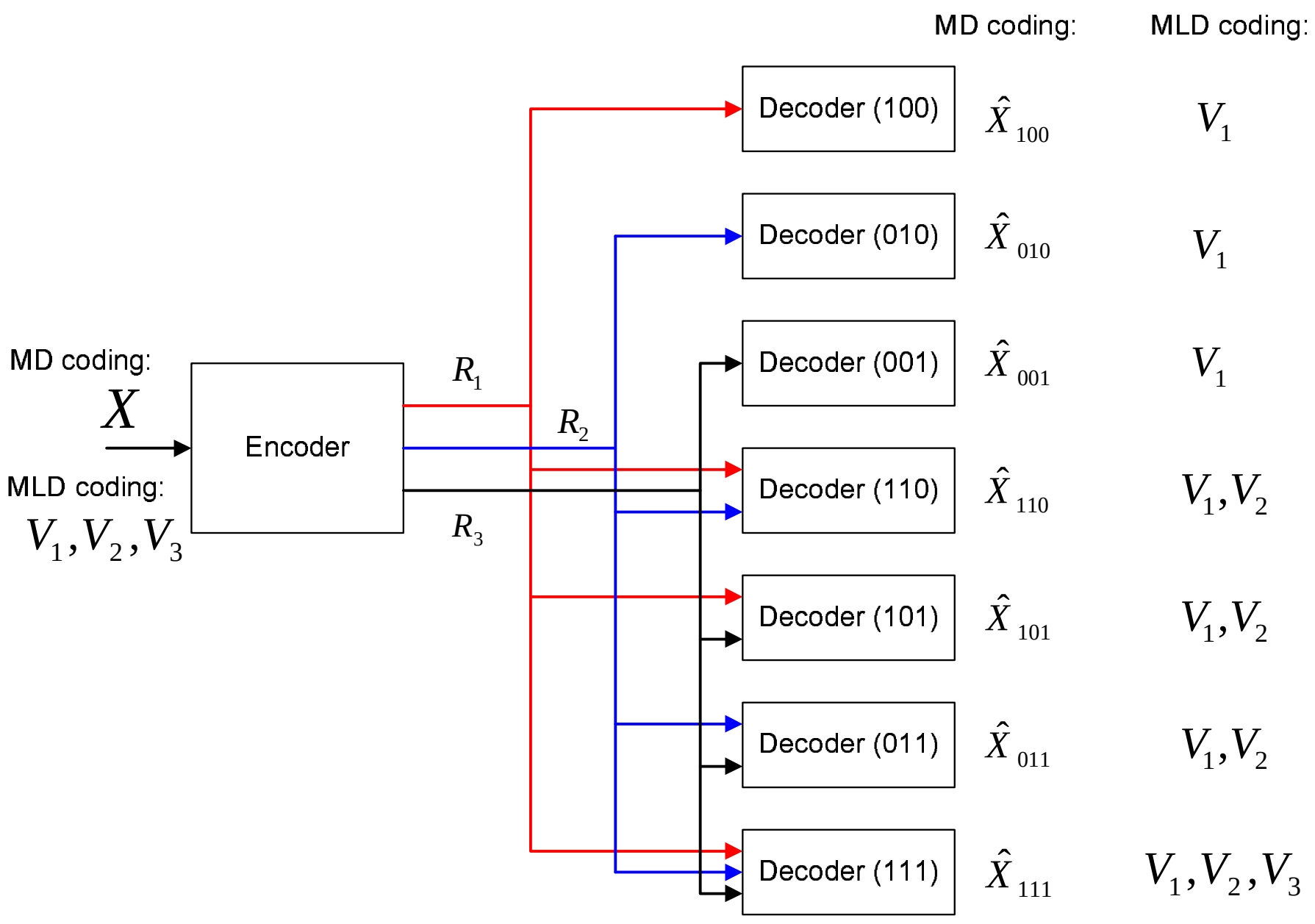

Figure 1: MD and MLD coding system diagrams for $K = 3$. More details on MLD coding are given in the next section.

The generalization of the second scheme of [5][6] has two aspects: we first show that the definition of the symmetric distribution, over which the scheme is optimized, can be relaxed straightforwardly; secondly by introducing additional coding component and invoking results on $\alpha$-resolution, we establish an achievable region that matches the polytopic template of MLD coding rate region. Interesting, the achievable rate region under a fixed set of auxiliary random variables is not a contra-polymatroid, unlike those often seen in other multiterminal source coding problems.

With the inner and outer bounds, we quantify the difference between them. For the symmetric rate problem, the individual-description rate-distortion (R-D) function can be bounded within a constant depending only on the number of descriptions, but not the distortion constraints. Moreover, regardless of the number of descriptions, the gap between the lower bound and the upper bound using the SR-ULP coding scheme is less than $1.48$ bits, and for the PPR multilayer scheme, the gap is less than $0.92$ bit. In order to establish these results, method similar to the enhancement technique in [19] is used. We also generalize the results to other sources under the mean squared error constraints, and show the sum rate gap between lower and upper bounds can be bounded within a constant, depending also only on the number of descriptions.

In addition to providing an approximate characterization of the symmetric individual-description R-D function, we also consider the *rate region* under symmetric distortion constraints. We first illustrate the basic ideas explicitly by considering the three-description case, and then extend the result to the general $K$-description problem. For the three-description case, we show that the outer and inner bounds can be repre-sented by ten planes with matching normal direction, and the Euclidean distances between the corresponding planes are shown to be less than certain small constants; these results are illustrated in Fig. 2. Then using the $\alpha$-resolution approach introduced in [16], we show that for the general $K$-description Gaussian problem under symmetric distortion constraints, the bounding planes of the rate region can be bounded both from above and below, between which the gap is bounded, and subsequently provide an approximate characterization of the R-D region.

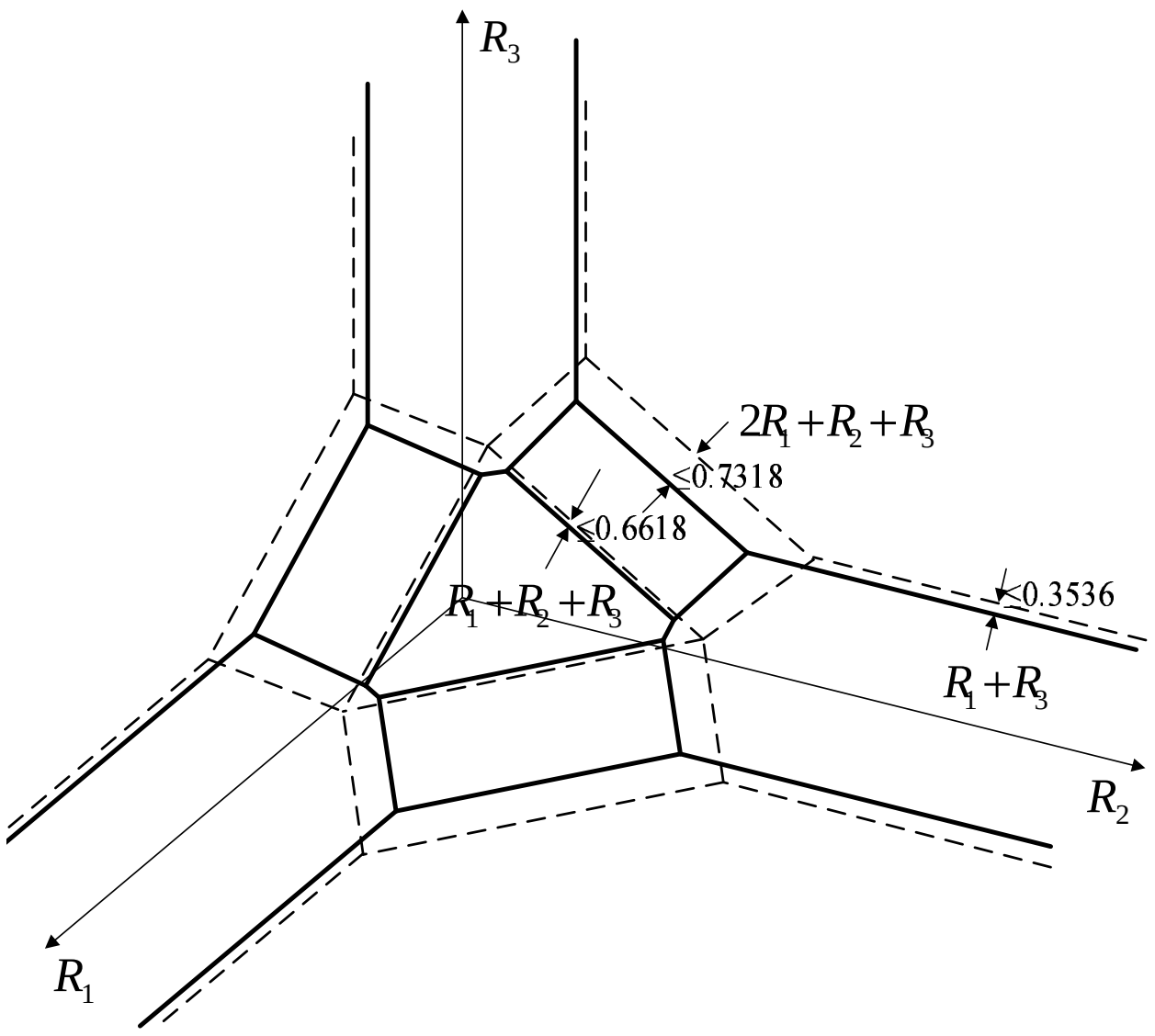

Figure 2: Bounding the rate distortion region for the three-description case, where the distances between corresponding planes of the inner and outer bounds are measured by Euclidean distance. The inner bound is drawn in with dashed lines, and the outer bound with solid lines.

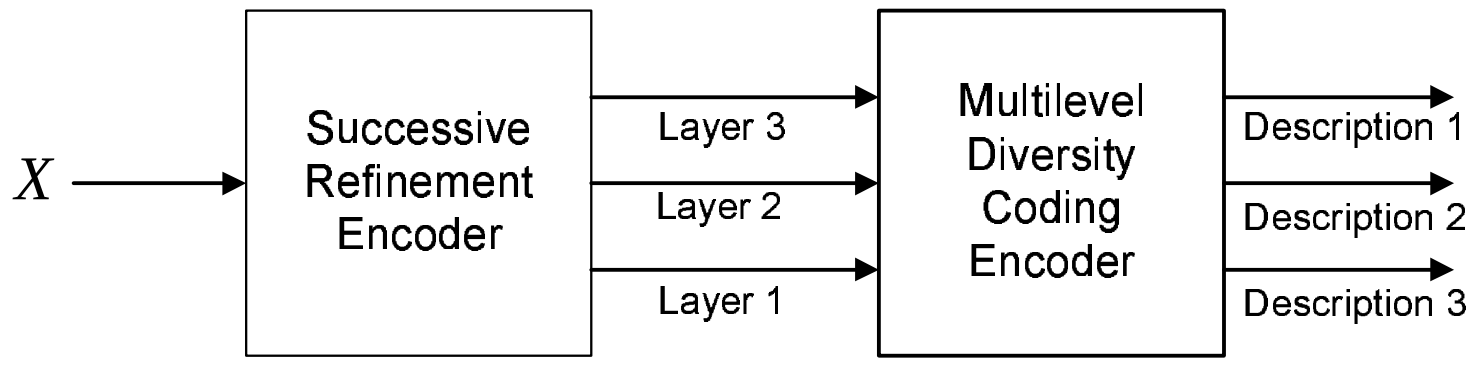

Figure 3: The separation approach based on successive refinement and lossless multi-level diversity coding.

It is surprising that the simple separation-based scheme of combining successive refinement and lossless multi-level diversity coding is able to achieve performance only a constant away from the optimal scheme; see Fig. 3 for the illustration of this system. This result implies that in certain practical high rate applications, this simple scheme may be sufficient, since additional gain will require much more complicated system design, and the resulting system will be significantly less flexible. Moreover, when distortion constraints are placed only on the last $k$ levels for the decoders with $K-k+1, K-k+2, ..., K$ descriptions, we show that even the gap between the lower and upper bound on the sum rate is asymptotically diminishing when the total number of descriptions $K$ becomes large with $k$ fixed. Thus virtually no gain is possible even in terms of sum rate for this case.

We emphasize that the general approach used in approximating the MD rate-distortion region is likely to provide insightful result for other network source (and channel) coding problems. More precisely, even though the exact rate-distortion region (or capacity region) of a multiuser information theory problem may have a general convex shape with a curvy boundary, simple polytopic inner and outer bounds are likely to exist which can provide a good approximate characterization. Comparing to general bounds, polytopic bounds are much easier to analyze. To apply this approach, it is desirable that the inner and outer bounds both follow a "common template" such that they can be conveniently compared. The result in our work suggests that a good choice of the template for a rate-distortion problem is the underlying lossless compression problem.

The rest of the paper is organized as follows. In Section 2 we provide a formal definition of the problem, and then briefly review the multi-level diversity problem and the $\alpha$-resolution method. In Section 3, we present a set of simplified results for the case with three descriptions as an illustrative example. Section 4 summarizes the main results of the paper. In Section 5, we focus on deriving the upper and lower bounds for the sum rate, and in Section 6, the inner and outer bounds for the rate region are presented. Finally Section 7 concludes the paper. Detailed and technical proofs are given in the appendices.

# 2 Notation, problem formulation and review

In this section we first provide the necessary notations and the problem definition, then briefly review the multi-level diversity coding problem and some essential $\alpha$-resolution results [15, 16] which play an important role in the development of our results. Wherever the notations or definitions become less transparent, we will specialize them to the three description case, i.e., the case $K = 3$. This special case will continue to serve as our working example, particularly in Section 3.

## 2.1 Notation and problem definition

Let $\{X(i)\}_{i=1,2,...}$ be a memoryless stationary source. At each time index $i$, the random variable $X(i)$ in an alphabet $\mathcal{X}$ is governed by the same distribution law $\mu_X$. In most of this work, we assume $\mathcal{X} = \mathbb{R}$, i.e., the real alphabet; moreover the reconstruction alphabet is also usually assumed to be $\mathbb{R}$. We use $\mathbb{R}_+$ to denote the set of non-negative reals. The vector $X(1), X(2), ..., X(n)$ will be denoted as $X^n$. Capital letters are used for random variables, and the corresponding lower-case letters are used for the realization of these random variables. Let $d : \mathcal{X} \times \mathcal{X} \to [0, \infty]$ be a single-letter distortion measure, and the multi-letter extension is defined as

$$d(x^n, y^n) = \frac{1}{n} \sum_{i=1}^{n} d(x(i), y(i)). \tag{1}$$

In this work, we are particularly interested in the squared error distortion measure $d(x, y) = (x-y)^2$. As such, it will be assumed without loss of generality that the source has a normalized unit variance. In this context, the most important case is the zero-mean unit-variance Gaussian source $X \sim \mathcal{N}(0, 1)$. In fact for the majority of this work we shall only consider the Gaussian source, except stated otherwise explicitly.

We shall adopt most of the notations in [16] introduced for the multi-level diversity coding (MLD) problem, which can be understood as a special case of the multiple description problem as we shall explain shortly. Throughout the paper, boldface letters are used to denote $K$-vectors. For the general $K$-description problem being considered, a length-$n$ block of the source samples is encoded into $K$ descriptions. Let $\boldsymbol{v}$ be a vector in $\{0, 1\}^K$, and denote the $i$-th component of $\boldsymbol{v}$ by $v_i$. Define

$$\Omega_K^\alpha = \{\boldsymbol{v} \in \{0, 1\}^K : |\boldsymbol{v}| = \alpha\}, \quad \alpha = 1, 2, ..., K \tag{2}$$

where $|\boldsymbol{v}|$ is the Hamming weight of $\boldsymbol{v}$, and define $\Omega_K = \bigcup_{\alpha=1}^{K} \Omega_K^\alpha$. Essentially, the set $\Omega_K$ provides a compact way to enumerate the possible combinations of the descriptions, or equivalently a compact way to enumerate the possible decoders. Particularly for the case of $K = 3$, we have

$$\Omega_3 = \Omega_3^1 \cup \Omega_3^2 \cup \Omega_3^3 = \{100, 010, 001\} \cup \{110, 101, 011\} \cup \{111\}. \tag{3}$$

Decoder $\boldsymbol{v}$, $\boldsymbol{v} \in \Omega_K$ has access to the $|\boldsymbol{v}|$ descriptions in the set $G_{\boldsymbol{v}} = \{i : v_i = 1\}$. For the case $K = 3$, we have

$$G_{100} = \{1\},\ G_{010} = \{2\},\ G_{001} = \{3\},\ G_{110} = \{1, 2\},\ G_{101} = \{1, 3\},\ G_{011} = \{2, 3\},\ G_{111} = \{1, 2, 3\}. \tag{4}$$

The *symmetric* distortion constraints are given such that any decoder $\boldsymbol{v}$ can reconstruct the source to satisfy a certain distortion $D_{|\boldsymbol{v}|}$, i.e., the distortion constraint depends only on the number of descriptions the decoder has access to, but not the particular combination of descriptions.

Formally, the problem is defined as follows. An $(n, (M_i, i \in I_K), (\Delta_{\boldsymbol{v}}, \boldsymbol{v} \in \Omega_K))$ code, where $I_K = \{1, 2, ..., K\}$, is defined as

$$S_i : \mathcal{X}^n \to I_{M_i}, \quad i \in I_K \tag{5}$$

$$T_{\boldsymbol{v}} : \textstyle\prod_{i \in G_{\boldsymbol{v}}} I_{M_i} \to \mathcal{X}^n, \quad \boldsymbol{v} \in \Omega_K, \tag{6}$$

and

$$\Delta_{\boldsymbol{v}} = \mathbb{E} d(X^n, \hat{X}^n_{\boldsymbol{v}}), \quad \boldsymbol{v} \in \Omega_K, \tag{7}$$

where

$$\hat{X}^n_{\boldsymbol{v}} = T_{\boldsymbol{v}}(S_i(X^n), i \in G_{\boldsymbol{v}}), \tag{8}$$

and $\mathbb{E}$ is the expectation operator. For the case $K = 3$, we have three encoders $S_1(\cdot)$, $S_2(\cdot)$ and $S_3(\cdot)$, and seven decoders $T_{100}$, $T_{010}$, $T_{001}$, $T_{110}$, $T_{101}$, $T_{011}$ and $T_{111}$, each decoder being associated with a reconstructed source sequence $\hat{X}^n_{\boldsymbol{v}}$ and inducing an expected distortion $\Delta_{\boldsymbol{v}}$.

A $K$-tuple $(R_1, R_2, ..., R_K)$ is $(D_1, D_2, ..., D_K)$-admissible if for every $\epsilon > 0$, there exists for sufficiently large $n$ an $(n, (M_i, i \in I_K), (\Delta_{\boldsymbol{v}}, \boldsymbol{v} \in \Omega_K))$ code such that

$$\frac{1}{n} \log M_i \le R_i + \epsilon, \quad i \in I_K, \tag{9}$$

and

$$\Delta_{\boldsymbol{v}} \le D_{|\boldsymbol{v}|} + \epsilon, \quad \boldsymbol{v} \in \Omega_K. \tag{10}$$

Throughout the paper, we use logarithm of base 2, such that the rate is measured by bits. Let $\mathcal{R}(\boldsymbol{D})$ be the collection of all $\boldsymbol{D}$-admissible rate vector, and this is the region of interest in this work. In the following sections, we shall assume $1 \ge D_1 \ge D_2 \ge ... \ge D_K > 0$ without loss of generality. One important special case is when the rates of the all the descriptions are the same, i.e., $M_i = M$ for any $i \in I_K$. For this symmetric rate case, the *symmetric individual-description* rate distortion (R-D) function $R(\boldsymbol{D})$ is defined simply as

$$R(\boldsymbol{D}) = \inf_{\{R : R \ge R_i, (R_1, R_2, ..., R_K) \in \mathcal{R}(\boldsymbol{D})\}} R. \tag{11}$$

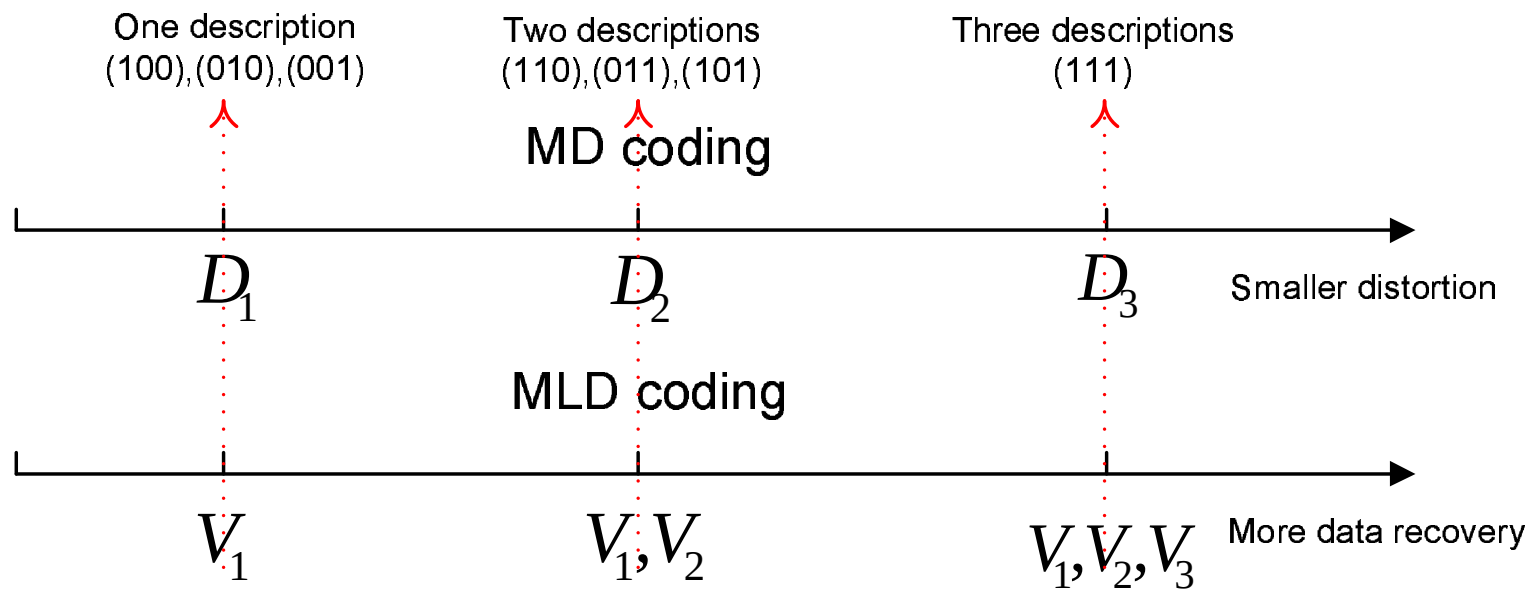

Figure 4: The similarity between MD coding and MLD coding for $K = 3$. They are essentially the same source coding problem with different distortion criteria.

Since $\mathcal{R}(\boldsymbol{D})$ is a closed set, the infimum can in fact be replaced by a minimum. Though in (11) we do not explicitly enforce the constraint that $R_1 = R_2 = ... = R_K$, it is straightforward to see this constraint can be added without causing any essential difference. For the case $K = 3$, we often expand the distortion vector and write the rate region and symmetric individual-description rate distortion function (SID-RD) as $\mathcal{R}(D_1, D_2, D_3)$ and $R(D_1, D_2, D_3)$, respectively.

Throughout the paper, when a rate $R$ is of interest, we use $\hat{R}$ or $\tilde{R}$ to denote its inner (upper) bounds, and use $\underline{R}$ to denote its outer (lower) bound; when rate region $\mathcal{R}$ is of interest, similar convention is taken.

## 2.2 A brief review of the symmetric multi-level diversity coding problem

The symmetric MLD coding problem considered in [15, 16] can be described as follows. A total of $K$ independent sources $V_1, V_2, ..., V_K$ are observed at the encoder, and encoded into $K$ descriptions. A decoder $T_{\boldsymbol{v}}$, which is called a level-$|\boldsymbol{v}|$ decoders, should reconstruct $V_1, V_2, ..., V_{|\boldsymbol{v}|}$ losslessly in the Shannon sense[1]. Particularly in the case of $K = 3$, three independent sources $V_1$, $V_2$ and $V_3$ are observed at the encoder, and encoded into three descriptions. The first level decoders $T_{100}$, $T_{010}$ and $T_{001}$ should reconstruct $V_1$ losslessly, the second level decoders $T_{110}$, $T_{101}$ and $T_{011}$ should reconstruct $(V_1, V_2)$, and the third level decoder $T_{111}$ should reconstruct $(V_1, V_2, V_3)$. The connection

In the framework of MD coding afore-introduced, we can simply treat the multi-source $V_1, V_2, ..., V_K$ as the single super source $X$, and the distortion measure $d_{|\boldsymbol{v}|}(\cdot, \cdot)$ is level-dependent, and thus also decoder-dependent, which is simply a Hamming distortion measure operating only on $V_1, V_2, ..., V_{|\boldsymbol{v}|}$. Therefore the lossless symmetric MLD coding problem essentially provides the solution to this symmetric MD problem at an extreme point of zero distortions for discrete memoryless sources; Fig. 1 and Fig. 4 illustrate the connection between the two problems in terms of the encoding/decoding functions and the distortion measure, respectively.

The main result for the symmetric MLD coding problem in [15, 16] is that source separation coding[2] is in fact optimal for this problem. The source separation coding scheme and the corresponding region can be described as follows. Each source vector $V_\alpha^n$ is encoded independently of the other sources, and the $i$-th description is allocated rate $r_i^\alpha$ for the $\alpha$-th source source $V_\alpha^n$. Each description is then the collection of encoded information (codes) produced for all the sources. The rate region is thus the set of non-negative rate

[1]It can be shown that lossless in the Shannon sense and lossless with diminishing Hamming distortion does not cause essential difference.

[2]This coding scheme was originally called superposition coding, but here we adopt the name *source separation coding* as suggested by Raymond Yeung, in order to avoid confusion with the superposition coding in broadcast channel.

vectors $\boldsymbol{R}$ that satisfy the following condition [16]

$$R_i = \sum_{\alpha=1}^{K} r_i^\alpha, \quad i = 1, 2, ..., K, \tag{12}$$

for some $r_i^\alpha \geq 0$, $\alpha = 1, 2, ..., K$ such that

$$\sum_{i \in G_{\boldsymbol{v}}} r_i^{|\boldsymbol{v}|} \geq H(V_{|\boldsymbol{v}|}), \quad \boldsymbol{v} \in \Omega_K. \tag{13}$$

The collection of information in all the descriptions may be redundant for any given source $V_\alpha$, $\alpha < K$, though any given specific description is maximumly compressed by itself. Clearly, the equality in (12) can be replaced by $\geq$ without loss of generality. As pointed out in [17], the condition (13) has an interpretation closely related to Slepian-Wolf coding, that the source words are randomly binned (for the $i$-th description) with rate $r_i^\alpha$, such that the source vector $V_\alpha^n$ can be recovered as long as the sum rate from any $\alpha$ descriptions for this source is larger than $H(V_\alpha)$. In [17], a connection to the maximum distance separable (MDS) codes was used to prove this result. Indeed, the Slepian-Wolf interpretation and the MDS codes interpretation are in fact closely related in this setting.

## 2.3 Review of the $\alpha$-resolution results

The rate region characterization (12) and (13) for MLD coding problem is given in a parametrized form, i.e., involving variables more than the rate tuple of interset $(R_1, R_2, .., .R_K)$. Though for smaller value of $K$, e.g., $K = 3$, it is possible to explicitly investigate the faces and vertex points of the rate region, for larger value of $K$ this becomes intractable. To overcome this difficulty, the $\alpha$-resolution method was invented in [16] to reveal the inherent structure of the MLD coding rate region. Next we directly quote a few definitions and results from [16]; some further results will be given after related notations are properly introduced. The readers in their initial reading may skip the lemmas and theorem in this subsection, and they will not be needed until Section 6.

Let $\boldsymbol{u}$ and $\boldsymbol{v}$ be two vectors in $\mathbb{R}^K$. Define $\boldsymbol{u} \geq \boldsymbol{v}$ if and only if $u_i \geq v_i$, $\forall i \in I_K$. Similar notation holds for $\boldsymbol{u}, \boldsymbol{v} \in \{0, 1\}^K$. For any $\boldsymbol{A} = (A_1, A_2, ..., A_K) \geq 0$, a mapping $c_\alpha : \Omega_K^\alpha \to \mathbb{R}_+$, where $\mathbb{R}_+$ is the set of non-negative real numbers, satisfying the following properties

$$c_\alpha(\boldsymbol{v}) \geq 0, \quad \text{for all} \quad \boldsymbol{v} \in \Omega_K^\alpha, \tag{14}$$

and

$$\sum_{\boldsymbol{v} \in \Omega_K^\alpha} c_\alpha(\boldsymbol{v})\boldsymbol{v} \leq \boldsymbol{A} \tag{15}$$

is called an $\alpha$-resolution for $\boldsymbol{A}$; it will be denoted as $\{c_\alpha(\boldsymbol{v})\}$ or simply as $c_\alpha$. Define a function $f_\alpha : \mathbb{R}_+^K \to \mathbb{R}_+$ for $\alpha \in I_K$ by

$$f_\alpha(\boldsymbol{A}) = \max \sum_{\boldsymbol{v} \in \Omega_K^\alpha} c_\alpha(\boldsymbol{v}), \tag{16}$$

where the maximum is taken over all the $\alpha$-resolution of $\boldsymbol{A}$. If $\{c_\alpha(\boldsymbol{v})\}$ achieves $f_\alpha(\boldsymbol{A})$, then it is called an optimal $\alpha$-resolution for $\boldsymbol{A}$, or simply $\alpha$-optimal. Without loss of generality up to a permutation of the rate vector components, we may assume

$$A_1 \geq A_2 \geq ... \geq A_K. \tag{17}$$

**Defintion 2.1** *Let $\{c_\alpha(\boldsymbol{v})\}$ be an $\alpha$-resolution of $\boldsymbol{A}$, then $\sum_{\boldsymbol{v}\in\Omega_K^\alpha} c_\alpha(\boldsymbol{v})\boldsymbol{v}$ is called the profile of $\{c_\alpha(\boldsymbol{v})\}$.*

**Lemma 2.1 ([16], Lemma 1)** *Let $\{c_\alpha(\boldsymbol{v})\}$ be $\alpha$-optimal for $\boldsymbol{A}$, and let $(\breve{A}_1, \breve{A}_2, ..., \breve{A}_K)$ be its profile. If there exist $1 \leq i \leq K$ such that $A_i - \breve{A}_i > 0$, then $c_\alpha(\boldsymbol{v}) > 0$ implies $v_i = 1$.*

**Lemma 2.2 ([16], Lemma 2)** *Let $\{c_\alpha(\boldsymbol{v})\}$ be $\alpha$-optimal for $\boldsymbol{A}$, and let $(\breve{A}_1, \breve{A}_2, ..., \breve{A}_K)$ be its profile, then there exists $0 \leq l_\alpha \leq \alpha - 1$ such that $A_i - \breve{A}_i > 0$ if and only if $1 \leq i \leq l_\alpha$.*

**Defintion 2.2** *For $2 \leq \alpha \leq K$, let $c_\alpha$ and $c_{\alpha-1}$ be $\alpha$-optimal and $(\alpha-1)$-optimal for $\boldsymbol{A}$, respectively. Then $c_{\alpha-1}$ covers $c_\alpha$, denoted by $c_{\alpha-1} \succ c_\alpha$, if*

$$\sum_{\boldsymbol{u}\in\Omega_K^{\alpha-1}} c_{\alpha-1}(\boldsymbol{u})H(S_i, i \in G_{\boldsymbol{u}}) \geq \sum_{\boldsymbol{v}\in\Omega_K^{\alpha}} c_\alpha(\boldsymbol{v})H(S_i, i \in G_{\boldsymbol{v}}), \tag{18}$$

*for any $K$ jointly distributed random variable $S_1, S_2, ..., S_K$.*

The following lemma is straightforward with the above definitions.

**Lemma 2.3** *Let $c_{\alpha-1}$ and $c_\alpha$ be $(\alpha-1)$-optimal and $\alpha$-optimal, respectively. If $c_{\alpha-1} \succ c_\alpha$, then $(\alpha - 1)f_{\alpha-1}(\boldsymbol{A}) \geq \alpha f_\alpha(\boldsymbol{A})$.*

**Proof 1 (Proof of Lemma 2.3)** *Let $S_1, S_2, ..., S_K$ be independently and identically distributed random variables with entropy $H(S_i) = H(S) > 0$ for any $i \in I_K$, then it follows*

$$\begin{aligned}(\alpha-1)f_{\alpha-1}(\boldsymbol{A})H(S) &= \sum_{\boldsymbol{u}\in\Omega_K^{\alpha-1}} c_{\alpha-1}(\boldsymbol{u})(\alpha-1)H(S) = \sum_{\boldsymbol{u}\in\Omega_K^{\alpha-1}} c_{\alpha-1}(\boldsymbol{u})H(S_i, i \in G_{\boldsymbol{u}}) \\ &\geq \sum_{\boldsymbol{v}\in\Omega_K^{\alpha}} c_\alpha(\boldsymbol{v})H(S_i, i \in G_{\boldsymbol{v}}) = \sum_{\boldsymbol{v}\in\Omega_K^{\alpha}} c_\alpha(\boldsymbol{v})\alpha H(S) = \alpha f_\alpha(\boldsymbol{A})H(S).\end{aligned} \tag{19}$$

*Dividing both ends by $H(S)$ completes the proof.*

By using Lemma 2.3 and the definition of $f_\alpha(\boldsymbol{A})$, the following lemma is rather immediate.

**Lemma 2.4** *The follows are true.*

- *The optimal 1-resolution is unique, $c_1(\boldsymbol{v}) = A_i$ for $G_{\boldsymbol{v}} = \{i\}$. Moreover $f_1(\boldsymbol{A}) = \sum_{k=1}^K A_i \triangleq A_{sum}$.*
- *The optimal $K$-resolution is unique $c_K(\boldsymbol{v}) = f_K(\boldsymbol{A}) = \min_{i\in I_K} A_i \triangleq A_{\min}$, where $G_{\boldsymbol{v}} = I_K$.*
- *For any $\alpha$ such that $K \geq \alpha \geq 2$, $f_\alpha(\boldsymbol{A}) \leq \frac{A_{sum}}{\alpha}$.*

The following theorem is instrumental for the result presented in [16], and it is also important for us to establish the result on the MD R-D region for $K > 3$.

**Theorem 2.1 ([16], Theorem 3)** *For any $\boldsymbol{A} \geq 0$, there exist $c_\alpha$, $1 \leq \alpha \leq K$, where $c_\alpha$ is $\alpha$-optimal for $\boldsymbol{A}$, such that*

$$c_1 \succ c_2 \succ ... \succ c_K. \tag{20}$$

# 3 A simple approximation for $K = 3$

In this section we give a set of approximate characterization of the SID-RD function and the rate-distortion region for the three description case. For the sake of simplicity, only the simple SR-MLD scheme is considered, and subsequently the set of results in this section is not as strong as those given in the following sections, however we choose to present them first for better exposition. The approximate rate-distortion region characterization for this case has a more explicit algebraic form, and can also be illustrated pictorially, which suites particularly well for the purpose of facilitating understanding. Moreover, as we shall show, even this set of simple results in fact provides a quite good approximation for the three description case.

## 3.1 Approximating the symmetric individual description rate distortion function

### 3.1.1 A simple upper bound

For the symmetric rate case, the source separation coding scheme reduces to the following simple unequal loss protection scheme; see, for example, [20]. Sources $V_1, V_2, V_3$ are losslessly compressed independent of each other. The encoded $V_1$ is repeated in all three description; a $(3, 2)$ maximum distance separable (MDS) code is applied to the encoded $V_2$ bitstream, and the resulting codeword is evenly split into each description; the encoded $V_3$ is then evenly split into each description without additional coding. This simple scheme clearly has the symmetric individual description rate of $H(V_1) + \frac{1}{2}H(V_2) + \frac{1}{3}H(V_3)$.

For the MD problem, consider now constructing the bitstream $B_i$ using the $i$-th layer of a successive refinement code for the Gaussian source, to satisfy the distortion constraint $D_i$, for $i = 1, 2, 3$. This coding structure is illustrated in Fig. 3, where the $i$-th layer output is taken to be the random source $V_i$. Since the quadratic Gaussian source is successively refinable [13], the following rate of $B_i$ is achievable

$$\hat{H}_i = \frac{1}{2}\log\frac{D_{i-1}}{D_i}, \quad i = 1, 2, 3. \tag{21}$$

where $D_0 \triangleq 1$.

With $B_i$ playing the role of the source vector $V_i^n$, it is clear that the following individual description rate is achievable, which provides a simple upper bound on the SID-RD function (defined in (11))

$$\hat{R}(D_1, D_2, D_3) = \frac{1}{2}\left[\log\frac{1}{D_1} + \frac{1}{2}\log\frac{D_1}{D_2} + \frac{1}{3}\log\frac{D_2}{D_3}\right] = \frac{1}{12}\log\frac{1}{D_1^3 D_2 D_3^2}. \tag{22}$$

### 3.1.2 A simple lower bound

Next we consider lower bounding the sum rate. To do this we write the following chain of inequalities.

$$\begin{aligned}
&n(R_1 + R_2 + R_3) \\
&\overset{(a)}{\geq} H(S_1) + H(S_2) + H(S_3) - H(S_1S_2S_3|X^n) \\
&\overset{(b)}{=} H(S_1) + H(S_2) + H(S_3) - H(S_1S_2S_3|X^n) - \frac{1}{2}[H(S_1S_2) + H(S_2S_3) + H(S_1S_3)] \\
&\quad + \frac{1}{2}[H(S_1S_2) + H(S_2S_3) + H(S_1S_3)] - H(S_1S_2S_3) + H(S_1S_2S_3) \triangleq \check{H}_3,
\end{aligned} \tag{23}$$

where (a) is because $S_i$, $i = 1, 2, 3$, are deterministic functions of $X^n$; (b) is by adding and subtracting the same term. This step may appear rather arbitrary, however a closer look reveals that the terms bear similarity to Han's inequality on subsets of random variables [22].

Next define $Y_2 = X + N_2$ and $Y_1 = X + N_1 + N_2$, where $N_1$ and $N_2$ are mutually independent Gaussian random variables, also independent of the Gaussian source $X$, with variance $\sigma_1^2$ and $\sigma_2^2$, respectively. Define $d_1 \triangleq \sigma_1^2 + \sigma_2^2$ and $d_2 \triangleq \sigma_2^2$, whose values are to be chosen later. The following step is essential for establishing the lower bound, which differs significantly from the technique of [2] and [8] in that we now utilize the two auxiliary random variables $Y_1$ and $Y_2$. Consider the following quantity

$$\acute{H}_3 = \left\{ H(S_1|Y_1^n) + H(S_2|Y_1^n) + H(S_3|Y_1^n) - \frac{1}{2}\left[H(S_1S_2|Y_1^n) + H(S_2S_3|Y_1^n) + H(S_1S_3|Y_1^n)\right] \right\} + \left\{ \frac{1}{2}\left[H(S_1S_2|Y_2^n) + H(S_2S_3|Y_2^n) + H(S_1S_3|Y_2^n)\right] - H(S_1S_2S_3|Y_2^n) \right\}. \tag{24}$$

It is seen that $\acute{H}_3 \geq 0$, because each brace in (24) is nonnegative by applying the conditional version of Han's inequality [22][3]. Intuitively, we expect certain conditional independence to hold approximately such that each brace is approximately zero. In this sense, the first brace roughly suggests that $Y_1$ is approximately a reconstruction with only (and any) one description, such that the individual descriptions are independent given $Y_1$; the second brace roughly suggests that $Y_2$ is approximately a reconstruction using only (and any) two descriptions, such that pairs of descriptions are independent given $Y_2$. Then it follows

$$\begin{aligned} n(R_1 + R_2 + R_3) &\geq \check{H}_3 - \acute{H}_3 \\ &= I(S_1; Y_1^n) + I(S_2; Y_1^n) + I(S_3; Y_1^n) + \frac{1}{2}[I(S_1S_2; Y_2^n) - I(S_1S_2; Y_1^n)] \\ &\quad + \frac{1}{2}[I(S_2S_3; Y_2^n) - I(S_2S_3; Y_1^n)] + \frac{1}{2}[I(S_1S_3; Y_2^n) - I(S_1S_3; Y_1^n)] \\ &\quad + [I(S_1S_2S_3; X^n) - I(S_1S_2S_3; Y_2^n)]. \end{aligned} \tag{25}$$

If $\acute{H}_3$ is close to zero, then the bounding above should yield meaningful result, which is indeed the case. We need the following lemma to proceed, the proof of which is in Appendix 8. Note that this lemma is not limited to the case of $K = 3$.

**Lemma 3.1** *Let $S_i$, $i \in I_K$ be a set of encoding functions such that there exist decoding functions to satisfy the distortion constraints $\boldsymbol{D} = (D_1, D_2, ..., D_K)$. Let $Y_b = X + N_b$ and $Y_a = X + N_a + N_b$, where $N_a$ and $N_b$ are mutually independent Gaussian random variables independent of the Gaussian source $X$, with variance $\sigma_a^2$ and $\sigma_b^2$, respectively. Then by defining $\sigma_b^2 = d_b$ and $\sigma_a^2 + \sigma_b^2 = d_a$, we have*

1. ***Mutual information bound between encoding functions and a noisy source***

$$I(S_i, i \in G_{\boldsymbol{v}}; Y_a^n) \geq \frac{n}{2} \log \frac{1 + d_a}{D_{|\boldsymbol{v}|} + d_a}, \tag{26}$$

2. ***Bound on mutual information different between encoding functions and different noisy sources***

$$I(S_i, i \in G_{\boldsymbol{v}}; Y_b^n) - I(S_i, i \in G_{\boldsymbol{v}}; Y_a^n) \geq \frac{n}{2} \log \frac{(1 + d_b)(D_{|\boldsymbol{v}|} + d_a)}{(1 + d_a)(D_{|\boldsymbol{v}|} + d_b)}. \tag{27}$$

[3]One can also optimize the distribution of auxiliary random variables $Y_1$ and $Y_2$, however in this work we only consider the specific Gaussian distribution given above, which yields relatively simple and easily computable bounds.

Clearly we can now apply the first statement in Lemma 3.1 to the first three terms in (25), and the second statement in Lemma 3.1 to the first three brackets in (25), by choosing appropriate $d_a$ and $d_b$. For the last bracket, let $\sigma_b^2 = 0$ and $\sigma_a^2 = \sigma_2^2$ in Lemma 3.1, then again the second statement can be applied.

Any valid choice of $d_1$ and $d_2$, i.e., $d_1 \geq d_2 > 0$, yields a valid lower bound. One could optimize within this set of lower bounds to find the tightest one, however without a matching inner bound, solving this rather involved optimization problem offers little insight. Instead, we shall choose some specific values, which indeed provides insightful results. Without loss of generality we may assume $D_1 \geq D_2 \geq D_3$. Thus $d_1 = D_1$ and $d_2 = D_2$ are a valid choice, and subsequently we have

$$\begin{aligned} R(D_1, D_2, D_3) &\geq \frac{1}{3}(R_1 + R_2 + R_3) \\ &\geq \frac{1}{12} \log \frac{(1+D_1)^3(1+D_2)(D_1+D_2)^3(D_2+D_3)^2}{2^9 D_1^6 D_2^3 D_3^2} \\ &\overset{(a)}{\geq} \frac{1}{12} \log \frac{1}{D_1^3 D_2 D_3^2} - \frac{3}{4}, \end{aligned} \tag{28}$$

where (a) is by using the facts $1 + D_i \geq 1$ and $D_i + D_{i+1} \geq D_i$ for $i = 1, 2$.

### 3.1.3 Comparing the upper and lower bounds

Combining (22) and (28), we have

$$\frac{1}{12} \log \frac{1}{D_1^3 D_2 D_3^2} \geq R(D_1, D_2, D_3) \geq \frac{1}{12} \log \frac{1}{D_1^3 D_2 D_3^2} - \frac{3}{4}. \tag{29}$$

The beginning and the end of inequalities differ only by a constant $\frac{3}{4}$ bit, which provides an approximation for the SID-RD function. This result reveals that the simple SR-ULP scheme is surprisingly competitive, since it is within $\frac{3}{4}$ bit of the optimum performance.

## 3.2 Approximating the rate-distortion region

### 3.2.1 A simple inner bound

For $K = 3$, the symmetric MLD coding rate region given in (12) and (13) can be written explicitly in the following form by applying the Fourier-Motzkin elimination [21] (see also [15])

$$R_i \geq H_1, \quad i = 1, 2, 3, \tag{30}$$

$$R_i + R_j \geq 2H_1 + H_2, \quad i \neq j, \quad i, j \in \{1, 2, 3\}, \tag{31}$$

$$2R_i + R_j + R_k \geq 4H_1 + 2H_2 + H_3, \quad (i, j, k) \text{ is a permutation of } (1, 2, 3), \tag{32}$$

$$R_1 + R_2 + R_3 \geq 3H_1 + \frac{3}{2}H_2 + H_3. \tag{33}$$

where $H_i \triangleq H(V_i)$ for $i = 1, 2, 3$.

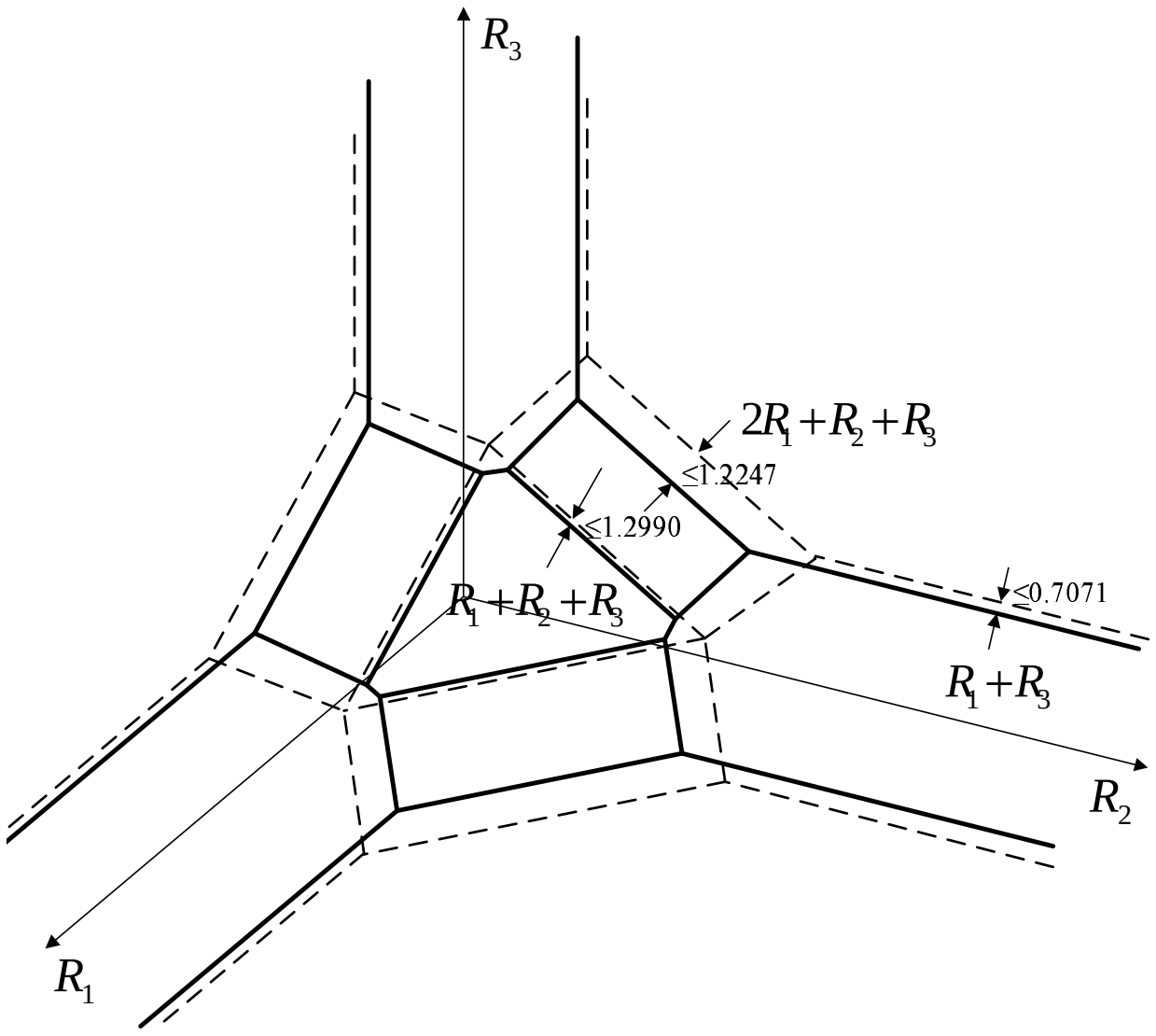

Figure 5: Simple inner and outer bounds for $\mathcal{R}(D_1, D_2, D_3)$. The gaps between the corresponding planes are measured by the Euclidean distance.

Clearly the achievability of the MLD coding rate region given by (30)-(33) implies that the following rate region is achievable for the MD problem by using the separation scheme illustrated in Fig. 3.

$$R_i \geq \frac{1}{2} \log \frac{1}{D_1}, \quad i = 1, 2, 3, \tag{34}$$

$$R_i + R_j \geq \frac{1}{2} \log \frac{1}{D_1 D_2}, \quad i \neq j, \quad i, j \in \{1, 2, 3\}, \tag{35}$$

$$2R_i + R_j + R_k \geq \frac{1}{2} \log \frac{1}{D_1^2 D_2 D_3}, \quad (i, j, k) \text{ is a permutation of } (1, 2, 3), \tag{36}$$

$$R_1 + R_2 + R_3 \geq \frac{1}{4} \log \frac{1}{D_1^3 D_2 D_3^2}. \tag{37}$$

#### 3.2.2 A simple outer bound

To derive an outer bound to match the template induced by the SR-MLD coding rate region, we need to consider bounding the rate combinations of $R_i$, $R_i + R_j$ and $2R_i + R_j + R_k$, in addition to the sum rate $R_1 + R_2 + R_3$. Clearly, the first two kinds of combination can be treated similarly as the sum rate, and we next show the last kind of rate combination can be appropriately bounded. We start with the following chain of inequalities,

$$\begin{aligned}
n(2R_i + R_j + R_k) &\geq 2H(S_i) + H(S_j) + H(S_k) \\
&\geq 2H(S_i) + H(S_j) + H(S_k) - H(S_i S_j) - H(S_i S_k) \\
&\quad + H(S_i S_j) + H(S_i S_k) - H(S_i S_j S_k) + H(S_i S_j S_k) \\
&\quad - [H(S_i|Y_1^n) + H(S_j|Y_1^n) - H(S_i S_j|Y_1^n)] - [H(S_i|Y_1^n) + H(S_k|Y_1^n) - H(S_i S_k|Y_1^n)] \\
&\quad - [H(S_i S_j|Y_2^n) + H(S_i S_k|Y_2^n) - H(S_i S_j S_k|Y_2^n)] - H(S_1 S_2 S_3|X^n),
\end{aligned} \tag{38}$$

where the brackets are nonnegative because $I(S_i;S_j|Y_1^n)$, $I(S_i;S_k|Y_1^n)$ and $I(S_iS_j;S_iS_k|Y_2^n)$ are nonnegative. Through some algebra, we arrive at

$$n(2R_i+R_j+R_k) \geq 2I(S_i;Y_1^n)+I(S_j;Y_1^n)+I(S_k;Y_1^n)+[I(S_iS_j;Y_2^n)-I(S_iS_j;Y_1^n)] \\ +[I(S_iS_k;Y_2^n)-I(S_iS_k;Y_1^n)]+[I(S_iS_jS_k;X^n)-I(S_iS_jS_k;Y_2^n)], \tag{39}$$

and now Lemma 3.1 can be applied. By taking $d_1 = D_1$ and $d_2 = D_2$ and further removing non-essential terms as in the sum rate case, an outer bound can be derived; the details are omitted here for brevity.

#### 3.2.3 Comparing the inner and outer bounds

With the simple inner and outer bounds, we conclude that the rate-distortion region is sandwiched between them as illustrated in Fig. 5, where the gaps between the corresponding planes are measured by the Euclidean distance. Note that the bounds given here are looser than those given in Fig. 2, and in later sections we will discuss how the tighter bounds are derived. In addition to providing an approximate characterization of the R-D region, the result further implies that the simple SR-MLD scheme is in fact not very far away from optimality, since it is within a small constant of the outer bound.

We use this section to illustrate the underlying ideas in the remainder of this paper. The result for the general $K$-description case given in the later sections are more involved, and we develop the general case result not only for the SR-MLD scheme, but also for the PPR multilayer scheme which is not separation-based. There are several difficulties in doing so: (1) There is no explicit representation of the inner and outer bounds as in the case for $K = 3$. (2) The PPR multilayer scheme is originally designed only for the symmetric-rate case, and we need to "inflate" the single rate point to a rate region. (3) To find tighter bounds, the simple choice for the values of $d_i$ used in this section is not sufficient. We first summarize the main results for the general $K$-description problem in Section 4, then in Section 5 and 6, we shall discuss in more details how these difficulties are addressed.

# 4 Main results

In this section, we present several theorems which summarize the main results for the Gaussian MD problem. The result on approximating the SID-RD function is first given, followed by the rate-distortion region approximation. More details are given in the Section 5, 6 and the appendices. Since the treatment for general sources under the MSE distortion measure is notationally more involved, they are thus delayed to those sections.

## 4.1 Approximating the symmetric individual description rate distortion function

Define the following functions

$$\hat{R}(\boldsymbol{D}) \triangleq \frac{1}{2}\sum_{\alpha=1}^{K}\frac{1}{\alpha}\log\frac{D_{\alpha-1}}{D_\alpha} \tag{40}$$

$$\tilde{R}(\boldsymbol{D}) \triangleq \frac{1}{2}\sum_{\alpha=1}^{K}\frac{1}{\alpha}\log\frac{D_{\alpha-1}}{D_\alpha} - \frac{1}{2}\sum_{\alpha=2}^{K}\frac{1}{\alpha}\left[\log\frac{\alpha-D_{\alpha-1}}{\alpha-1}\right] \tag{41}$$

$$\underline{R}(\boldsymbol{D},\boldsymbol{d}) \triangleq \frac{1}{2}\sum_{\alpha=1}^{K}\frac{1}{\alpha}\log\frac{(1+d_\alpha)(D_\alpha+d_{\alpha-1})}{(1+d_{\alpha-1})(D_\alpha+d_\alpha)}, \tag{42}$$

where $d_1 \geq d_2 \geq ... \geq d_{K-1} > 0$, $d_0 \triangleq \infty$ and $d_K \triangleq 0$; $D_0 \triangleq 1$ and we take the convention $\log \frac{\infty}{\infty} = 0$. For convenience we define

$$\underline{R}(\boldsymbol{D}) \triangleq \sup_{d_1 \geq d_2 \geq ... \geq d_{K-1} > 0} \underline{R}(\boldsymbol{D}, \boldsymbol{d}), \tag{43}$$

Define the following functions

$$\Phi_\alpha(D) = \frac{\alpha D}{1-D}, \quad \alpha = 1, 2, ..., K. \tag{44}$$

For a given distortion vector $\boldsymbol{D} = (D_1, D_2, ..., D_K)$, we shall associate it with an *enhanced distortion vector* $\boldsymbol{D^*} = (D_1^*, D_2^*, ..., D_K^*)$ using a recursive procedure.

$$\begin{aligned} D_1^* &= D_1, \\ D_k^* &= \begin{cases} \frac{(\alpha-1)D_{\alpha-1}^*}{\alpha - D_{\alpha-1}^*} & \Phi_\alpha(D_\alpha) > \Phi_{\alpha-1}(D_{\alpha-1}^*) \\ D_k & \text{otherwise} \end{cases}, \quad k = 2, 3, ..., K. \end{aligned} \tag{45}$$

This enhanced distortion vector is introduced in order to remove certain cases where the given distortion vectors can not be satisfied with equality using the coding schemes we consider; moreover, it has the property that it does not significantly effect the lower bound. More details on the enhanced distortion vector are given in Section 5.B. We shall also assume $D_1 < 1$ for simplicity at this point, but will discuss the cases when $D_1 = 1$ shortly.

We are now ready to present the main theorem of this subsection.

**Theorem 4.1** *Let $\boldsymbol{D^*}$ be the enhanced distortion vector of $\boldsymbol{D}$, then the Gaussian SID-RD function under symmetric distortion constraints satisfies*

$$\hat{R}(\boldsymbol{D^*}) \geq \tilde{R}(\boldsymbol{D^*}) \geq R(\boldsymbol{D}) \geq \underline{R}(\boldsymbol{D}) \geq \underline{R}(\boldsymbol{D}, \boldsymbol{d}), \tag{46}$$

*for any $d_1 \geq d_2 \geq ... \geq d_{K-1} > d_K = 0$ and $d_0 \triangleq \infty$. Moreover,*

$$\begin{aligned} \hat{R}(\boldsymbol{D^*}) - \underline{R}(\boldsymbol{D}) &\leq \frac{1}{2}\sum_{\alpha=2}^{K} \frac{1}{\alpha-1} \log \alpha - \frac{1}{2}\sum_{\alpha=2}^{K} \frac{1}{\alpha} \log(\alpha - 1) \triangleq \hat{L}(K) \leq 1.48, & (47) \\ \tilde{R}(\boldsymbol{D^*}) - \underline{R}(\boldsymbol{D}) &\leq \frac{1}{2}\sum_{\alpha=2}^{K} \left[\frac{1}{\alpha-1} - \frac{1}{\alpha}\right] \log \alpha \triangleq \tilde{L}(K) \leq 0.92. & (48) \end{aligned}$$

*Remark:* In Theorem 4.1, we bound the gaps between the inner and outer bounds by universal constants. This is not necessary, and we will show in Section 5 that the bounds can in fact be distortion dependent, however we relax these bounds to make it universal here. The numerical values are derived using integral approximation for series which does not yield the tightest bounds possible. In Table 1 we have included a few values of these bounds.

An important and interesting special case is when only the last several levels have distortion constraints, since usually the packet loss probability is not exceedingly high, and for the majority of the time only a small number of packets can be lost. Though the universal bound in Theorem 4.1 also holds for degenerate cases where only certain levels of distortion constraints exist, applying the theorem using the general bound $\underline{R}(\boldsymbol{D}, \boldsymbol{d})$ can improve the universal constants significantly. In order to do so, the values $(d_1, d_2, ..., d_{K-1})$ need to be chosen carefully.

Table 1: Values $\hat{L}(K)$ and $\tilde{L}(K)$ for $K = 1, 2, 3, ..., 8$.

| $K$ | 2 | 3 | 4 | 5 | 6 | 7 | 8 |
|---|---|---|---|---|---|---|---|
| $\hat{L}(K)$ | 0.5000 | 0.7296 | 0.8648 | 0.9550 | 1.0200 | 1.0693 | 1.1082 |
| $\tilde{L}(K)$ | 0.2500 | 0.3821 | 0.4654 | 0.5235 | 0.5665 | 0.6000 | 0.6268 |

**Corollary 4.1** *For the Gaussian source, when only distortion constraints $D_{K-k+1}, D_{K-k+2}, ..., D_K$ exist for $k \in I_K$, (or equivalently $D_1 = D_2 = ... = D_{K-k} = 1$,) we have*

$$\hat{R}(\boldsymbol{D}^*) - \underline{R}(\boldsymbol{D}) \leq \frac{1}{2}\sum_{\alpha=K-k+2}^{K}\frac{1}{\alpha-1}\log\alpha - \frac{1}{2}\sum_{\alpha=K-k+2}^{K}\frac{1}{\alpha}\log(\alpha-1)$$

$$\tilde{R}(\boldsymbol{D}^*) - \underline{R}(\boldsymbol{D}) \leq \frac{1}{2}\sum_{\alpha=K-k+2}^{K}\left[\frac{1}{\alpha-1} - \frac{1}{\alpha}\right]\log\alpha. \tag{49}$$

*Remark:* These bounds are usually significantly tighter than the constants given in Theorem 4.1. It is easily seen that when $k$ is kept fixed and $K \to \infty$, the gap approaches zero; in fact, in this case even the sum rate bounds become asymptotically tight. Corollary 4.1 thus implies that the SR-ULP scheme is even more closer to optimum, and the benefit of using more complicated schemes is diminishing as the number of description increases, when we are guaranteed to receive all but a constant number of descriptions.

## 4.2 Approximating the rate-distortion region

We first define two regions, which are in fact two inner bounds to the Gaussian MD rate region. The first region is based on the SR-MLD scheme illustrated in Fig. 3, and now we define this (achievable) region for general $K > 3$. Let $\hat{\mathcal{R}}(\boldsymbol{D})$ be the set of non-negative rate vectors $(R_1, R_2, ..., R_K)$, such that

$$R_i \geq \sum_{\alpha=1}^{K} r_i^{\alpha}, \quad 1 \leq i \leq K, \tag{50}$$

for some $r_i^{\alpha} \geq 0$, $1 \leq \alpha \leq K$, satisfying

$$\sum_{i \in G_{\boldsymbol{v}}} r_i^{|\boldsymbol{v}|} \geq \hat{H}_{|\boldsymbol{v}|}(\boldsymbol{D}), \quad \forall \boldsymbol{v} \in \Omega_K, \tag{51}$$

where

$$\hat{H}_{\alpha}(\boldsymbol{D}) = \frac{1}{2}\log\frac{D_{\alpha-1}}{D_{\alpha}}, \quad \alpha = 1, 2, ..., K, \tag{52}$$

and $D_0 \triangleq 1$. It is clearly that since the Gaussian source is successively refinable, the right hand side of (52) simply gives the rate for each layer in the optimal successive refinement code; (50) and (51) are simply the counterpart of (12) and (13).

The second region is based on a generalization of the PPR multilayer scheme, the details of which are given in Section 6. First let $\boldsymbol{D}^*$ be the enhanced distortion vector of $\boldsymbol{D}$ and define the following quantities

$$\begin{aligned}\tilde{H}_1(\boldsymbol{D}^*) &= \frac{1}{2}\log\frac{1}{D_1^*},\\ \tilde{H}_\alpha(\boldsymbol{D}^*) &= \frac{1}{2}\log\frac{(\alpha-1)D_{\alpha-1}^*}{(\alpha-D_{\alpha-1}^*)D_\alpha^*},\quad \alpha=2,3,...,K.\end{aligned} \tag{53}$$

Let $\tilde{\mathcal{R}}(\boldsymbol{D}^*)$ be the set of non-negative rate vectors $(R_1, R_2, ..., R_K)$, such that

$$R_i \geq \sum_{\alpha=1}^{K} r_i^\alpha,\quad 1\leq i\leq K, \tag{54}$$

for some $r_i^\alpha \geq 0$, $1\leq \alpha\leq K$, satisfying

$$\sum_{i\in G_{\boldsymbol{v}}} r_i^{|\boldsymbol{v}|} \geq \tilde{H}_{|\boldsymbol{v}|}(\boldsymbol{D}^*),\quad \boldsymbol{v}\in\Omega_K. \tag{55}$$

The following theorem establishes that both $\hat{\mathcal{R}}(\boldsymbol{D})$ and $\tilde{\mathcal{R}}(\boldsymbol{D})$ are inner bounds to the Gaussian MD rate-distortion region.

**Theorem 4.2** *Let $\boldsymbol{D}^*$ be the enhanced distortion vector of $\boldsymbol{D}$,*

$$\hat{\mathcal{R}}(\boldsymbol{D}^*) \subseteq \tilde{\mathcal{R}}(\boldsymbol{D}^*) \subseteq \mathcal{R}(\boldsymbol{D}^*) \subseteq \mathcal{R}(\boldsymbol{D}). \tag{56}$$

For $K>3$, it is difficult to enumerate the faces of the inner and outer bounds, thus we alternatively seek to approximately characterize the bounding planes of the rate-distortion region, defined for any $\boldsymbol{A}\in\mathbb{R}_+^K$ and $\boldsymbol{A}\neq 0$, as the following function

$$R_{\boldsymbol{A}}(\boldsymbol{D}) \triangleq \min_{\boldsymbol{R}\in\mathcal{R}(\boldsymbol{D})} \boldsymbol{A}\cdot\boldsymbol{R} \tag{57}$$

Define the following function

$$\underline{R}_{\boldsymbol{A}}(\boldsymbol{D},\boldsymbol{d}) \triangleq \frac{1}{2}\sum_{\alpha=1}^{K} f_\alpha(\boldsymbol{A})\log\frac{(1+d_\alpha)(D_\alpha+d_{\alpha-1})}{(1+d_{\alpha-1})(D_\alpha+d_\alpha)}. \tag{58}$$

where the function $f_\alpha(\boldsymbol{A})$ is defined in (16) and $d_1\geq d_2\geq ...\geq d_{K-1}>0$, $d_0\triangleq\infty$ and $d_K\triangleq 0$. Define further the following function

$$\underline{R}_{\boldsymbol{A}}(\boldsymbol{D}) \triangleq \sup_{d_1\geq d_2\geq...\geq d_{K-1}>0} \underline{R}_{\boldsymbol{A}}(\boldsymbol{D},\boldsymbol{d}). \tag{59}$$

The next theorem establishes the upper and lower bounds for the bounding planes of the rate-distortion region. Since the rate-distortion region is convex, if the upper and lower bounds for the bounding planes coincide, a complete characterization is then available. The upper and lower bounds given in the following theorem do not coincide in general, however the gap between them is bounded, yielding an approximate characterization of the rate region.

Table 2: The values of $f_\alpha(\boldsymbol{A})$ and bounds for $K=3$.

| $\boldsymbol{A}$ | $(1,0,0)$ | $(1,1,0)$ | $(2,1,1)$ | $(1,1,1)$ |
|---|---|---|---|---|
| $f_1(\boldsymbol{A})$ | 1 | 2 | 4 | 3 |
| $f_2(\boldsymbol{A})$ | 0 | 1 | 2 | 1.5 |
| $f_3(\boldsymbol{A})$ | 0 | 0 | 1 | 1 |
| $\\|\cdot\\|$ by (61) | 0 | $\frac{1}{\sqrt{2}}$ | $\frac{3+2\log 3}{2\sqrt{6}}$ | $\frac{4+3\log 3}{4\sqrt{3}}$ |
| $\\|\cdot\\|$ by (62) | 0 | $\frac{1}{2\sqrt{2}}$ | $\frac{2+\log 3}{2\sqrt{6}}$ | $\frac{3+\log 3}{4\sqrt{3}}$ |

**Theorem 4.3** *For the Gaussian source and any $\boldsymbol{A}\geq 0$,*

$$\textstyle\sum_{\alpha=1}^K f_\alpha(\boldsymbol{A})\hat{H}_\alpha(\boldsymbol{D}^*)\geq\sum_{\alpha=1}^K f_\alpha(\boldsymbol{A})\tilde{H}_\alpha(\boldsymbol{D}^*)\geq R_{\boldsymbol{A}}(\boldsymbol{D})\geq \underline{R}_{\boldsymbol{A}}(\boldsymbol{D})\geq \underline{R}_{\boldsymbol{A}}(\boldsymbol{D},\boldsymbol{d}) \tag{60}$$

*for any $d_1\geq d_2\geq ...\geq d_{K-1}>0$, $d_0\triangleq\infty$ and $d_K\triangleq 0$. Moreover, for any $\boldsymbol{A}\in\mathbb{R}_+^K$ and $\boldsymbol{A}\neq 0$,*

$$\begin{aligned}\sum_{\alpha=1}^K f_\alpha(\boldsymbol{A})\hat{H}_\alpha(\boldsymbol{D}^*)-\underline{R}_{\boldsymbol{A}}(\boldsymbol{D})&\leq\frac{1}{2}\sum_{\alpha=2}^K f_{\alpha-1}(\boldsymbol{A})\log\alpha-\frac{1}{2}\sum_{\alpha=2}^K f_\alpha(\boldsymbol{A})\log(\alpha-1)\\&\leq\frac{A_{sum}}{2}\sum_{\alpha=2}^K\frac{1}{\alpha-1}\log\alpha-\frac{A_{sum}}{2}\sum_{\alpha=2}^{K-1}\frac{1}{\alpha}\log(\alpha-1)-\frac{A_{\min}}{2}\log(K-1),\end{aligned}\tag{61}$$

*and*

$$\begin{aligned}\sum_{\alpha=1}^K f_\alpha(\boldsymbol{A})\tilde{H}_\alpha(\boldsymbol{D}^*)-\underline{R}_{\boldsymbol{A}}(\boldsymbol{D})&\leq\frac{1}{2}\sum_{\alpha=2}^K[f_{\alpha-1}(\boldsymbol{A})-f_\alpha(\boldsymbol{A})]\log\alpha\\&\leq\frac{A_{sum}}{2}\sum_{\alpha=2}^{K-1}[\frac{1}{\alpha-1}-\frac{1}{\alpha}]\log\alpha+\frac{1}{2}(\frac{A_{sum}}{K-1}-A_{\min})\log(K).\end{aligned}\tag{62}$$

*Remark:* It is not immediately clear that the outer bound, which is specified in terms of an uncountable number of bounding planes indexed by $\boldsymbol{A}$, is still a polytope as for the case $K=3$. Nevertheless it can indeed be shown that when we specialize these bounds for appropriate choice of $\boldsymbol{d}$, it is an equivalent characterization of a polytope. Moreover, the bounds given in (61) and (62) are established using the bound induced by this specific choice of $\boldsymbol{d}$. We shall return to this point with more details in Section 6.

*Remark:* Theorem 4.3, which provides approximate characterizations of the rate-distortion region, is given in a similar manner as Theorem 4.1, which provides approximation characterizations of the SID-RD function. The second bound in (61) and the second bound in (62) are more explicit, whereas the first bounds involve the function $f_\alpha(\boldsymbol{A})$ which requires solving an optimization problem. These bounds imply that the gaps between the bounding planes of inner and outer bounds is upper-bounded by constants independent of the distortion constraints.

*Remark:* Whether the polytopic inner bound is a good approximate characterization of the rate region does not depend on whether the outer bound is a polytope, but only on how large the gap is between the inner and outer bounds. Though for the Gaussian source, the outer bound can be specialized to be a polytope, for general sources this does not necessarily hold. Nevertheless, even for general sources, the inner bound, which is an approximate characterization of the rate region, is still a polytope.

**Example for** $K = 3$**:** Now we apply the result in Theorem 4.3 to the case of $K = 3$. As illustrated in Section 3, it suffices to consider the choices of vector $\boldsymbol{A}$ in the following set

$$\{(1,0,0),(1,1,0),(2,1,1),(1,1,1)\}. \tag{63}$$

In Table 2, we list the value of $f_\alpha(\boldsymbol{A})$, which can be easily verified since the $\alpha$-resolution formulation is a linear optimization problem. Using (61) and (62), it is straightforward to compute the bounds between the inner and upper bounds, as shown in the last two rows of Table 2. Note that here the distance is normalized in terms of Euclidean distance. This improves the result given in Section 3, which was illustrated in Fig. 2.

# 5 Sum rate bound and SID-RD function approximation

In this section, we provide more details on the derivation of results regarding SID-RD function. Some intermediate results will be given, which may in fact be of interest by themselves when tighter distortion dependent bounds are needed. We first introduce more formally two achievable individual description rates, which are given in a general form that can also be applied to other sources, then the derivation of the outer bounds is discussed. With both the inner and outer bounds, we analyze and bound the gap between them. Finally, we extend the results to general sources under the MSE distortion measure.

## 5.1 Achievable rate using the SR-ULP scheme

The SR-MLD coding scheme reduces to the SR-ULP scheme when the rate is also symmetric, i.e., $R_1 = R_2 = ... = R_K$. For a general source, we have the following theorem.

**Theorem 5.1** *For any given set of random variables* $(Y_1, Y_2, ..., Y_K)$ *jointly distributed with the source* $X$*, such that there exist deterministic functions* $g_\alpha : \mathcal{Y}^\alpha \to \mathcal{X}$ *to satisfy*

$$\mathbb{E}d(X, g_\alpha(Y_1, Y_2, ..., Y_\alpha)) \leq D_\alpha, \quad \alpha = 1, 2, ..., K, \tag{64}$$

*we have*

$$R(D_1, D_2, ..., D_K) \leq \sum_{\alpha=1}^{K} \frac{1}{\alpha} I(X; Y_\alpha | Y_1, Y_2, \ldots, Y_{\alpha-1}), \tag{65}$$

*where* $Y_0 \triangleq 0$.

This theorem is a natural consequence of combining the result on successive refinement [13, 14] and the property of the MDS codes, and thus the proof is omitted. This theorem is given formally in order to facilitate the analysis for general sources. In this work, we consider the following natural distribution often seen in the successive refinement problem

$$Y_\alpha = X + \sum_{i=\alpha}^{K} N_i, \quad \alpha = 1, 2, ..., K \tag{66}$$

where $N_i \sim \mathcal{N}(0, \sigma_i^2)$ are mutually independent and also independent of $X$. For convenience, we denote $\sum_{i=\alpha}^{K} N_i$ as $Z_\alpha$. The values of variance $\sigma_i^2$, $i = 1, 2, ..., K$ are chosen such that the distortion constraint at each level is satisfied with equality when the reconstruction is the linear minimum mean squared error estimator (LMMSE), i.e., they are determined by the set of equations

$$D_\alpha = \frac{\sum_{i=\alpha}^{K} \sigma_i^2}{1 + \sum_{i=\alpha}^{K} \sigma_i^2}, \quad \alpha = 1, 2, ..., K. \tag{67}$$

It is clear that there always exists a unique and valid solution for these variances when the distortion constraints are given in the natural monotonic order. Through basic algebraic calculation, we arrive at the following corollary for $\hat{R}(\boldsymbol{D})$ defined in (40),

**Corollary 5.1** *For the Gaussian source,*

$$R(\boldsymbol{D}) \leq \hat{R}(\boldsymbol{D}). \tag{68}$$

## 5.2 Achievable rate using the PPR multilayer scheme

In the two-part paper [5] and [6], an achievable symmetric individual rate is given for the symmetric MD problem, and the main theorem is quoted below together with a necessary definition.

**Theorem 5.2 ([6] Theorem 2)** *For any probability distribution*

$$p(x, \{y_{\alpha,j}, \alpha \in I_{K-1}, j \in I_K\}, y_K) = p(x)p(\{y_{\alpha,j}, \alpha \in I_{K-1}, j \in I_K\}, y_K|x), \tag{69}$$

*where* $p(\{y_{\alpha,j}, \alpha \in I_{K-1}, j \in I_K\}, y_K|x)$ *is symmetric over* $\mathcal{X} \times \mathcal{Y}^{K(K-1)+1}$ *and a set of decoding functions*

$$\begin{aligned} g_{\boldsymbol{v}} &: \mathcal{Y}^{|\boldsymbol{v}||\boldsymbol{v}|} \to \mathcal{X}, \quad \boldsymbol{v} \in \Omega_K,\ |\boldsymbol{v}| < K, \\ g_{\boldsymbol{v}} &: \mathcal{Y}^{K(K-1)+1} \to \mathcal{X}, \quad |\boldsymbol{v}| = K, \end{aligned} \tag{70}$$

*such that*

$$\begin{aligned} \mathbb{E}(d(X, g_{\boldsymbol{v}}(Y_{\alpha,j}, \alpha \in I_{|\boldsymbol{v}|}, j \in G_{\boldsymbol{v}}))) &\leq D_{|\boldsymbol{v}|}, \quad \boldsymbol{v} \in \Omega_K,\ |\boldsymbol{v}| < K, \\ \mathbb{E}(d(X, g_{\boldsymbol{v}}(\{Y_{\alpha,j}, \alpha \in I_{|\boldsymbol{v}|}, j \in G_{\boldsymbol{v}}\}, Y_K))) &\leq D_K, \quad |\boldsymbol{v}| = K, \end{aligned} \tag{71}$$

*the following symmetric individual description rate is achievable*

$$\begin{aligned} R &= \sum_{\alpha=1}^{K-1} \frac{1}{\alpha} H(Y_{\alpha,j}, j \in I_\alpha | Y_{i,j}, i \in I_{\alpha-1}, j \in I_\alpha) \\ &\quad + \frac{1}{K} H(Y_K | Y_{i,j}, i \in I_{K-1}, j \in I_K) - \frac{1}{K} H(\{Y_{i,j}, i \in I_{K-1}, j \in I_K\}, Y_K | X). \end{aligned} \tag{72}$$

A symmetric distribution is defined in [6] as follows.

**Defintion 5.1** *A joint distribution* $p(\{y_{\alpha,j}, \alpha \in I_{K-1}, j \in I_K\}, y_K|x)$ *is called symmetric if for all* $1 \leq n_i \leq K$ *where* $i \in I_{K-1}$*, the following is true: the joint distribution of* $Y_K$ *and all* $(n_1 + n_2 + ... + n_{K-1})$ *random variables where any* $n_\alpha$ *are chosen from the set* $\{Y_{\alpha,1}, Y_{\alpha,2}, ..., Y_{\alpha,K}\}$*, conditioned on* $X$*, is the same.*

Intuitively, the PPR multilayer scheme provides layered information in the descriptions, and the $\alpha$-th layer information can only be decoded when at least $\alpha$ descriptions are available. The encoding auxiliary random variable $Y_{\alpha,j}$ is essentially the information provided in the $j$-th description for the $\alpha$-th layer. In [5] and [6], a clever scheme of organizing the information is given, resulting in the achievable rate given in Theorem 5.2.

We notice that the definition in Definition 5.1 is however unnecessarily restrictive and can be straightforwardly relaxed. The following alternative definition of a symmetric distribution can replace the more restrictive one. This relaxed version of symmetric distribution will be useful since our choice of distribution, which provides simplification in computing the inner bound, is in this relaxed set, but not in the original more restrictive set.

**Defintion 5.2** *A joint distribution $p(\{y_{\alpha,j}, \alpha \in I_{K-1}, j \in I_K\}, y_K|x)$ is called* generalized symmetric *if for any permutation $\pi(\cdot) : I_K \to I_K$, the joint distribution $p(\{y_{\alpha,\pi(j)}, \alpha \in I_{K-1}, j \in I_K\}, y_K|x)$ is the same as $p(\{y_{\alpha,j}, \alpha \in I_{K-1}, j \in I_K\}, y_K|x)$.*

It is straightforward to check that Theorem 5.3 holds true, when we replace the requirement of symmetric distribution with the generalized version. The original version of symmetric distribution essentially requires the distribution to be invariant under $K-1$ different permutations $\pi_\alpha(\cdot)$, one for each layer $\alpha = 1, 2, ..., K-1$; i.e., if we permute $\{Y_{1,1}, Y_{1,2}, ..., Y_{1,K}\}$, and then permute $\{Y_{2,1}, Y_{2,2}, ..., Y_{3,K}\}$ differently, and so on for each $\alpha = 1, 2, ..., K-1$, the resulting distribution should remain the same as the one before such permutations. This requirement was however never completely utilized in the coding scheme. Instead the coding scheme in fact only requires invariance under a single permutation $\pi(\cdot)$ which is applied to all the levels simultaneously, i.e., $\pi_\alpha(\cdot) = \pi(\cdot)$, for $\alpha = 1, 2, ..., K-1$. More formally, we state this generalized result as a theorem.

**Theorem 5.3** *The statement of Theorem 5.2 holds when the symmetric distribution requirement is replaced with the generalized symmetric distributions.*

From Theorem 5.3, an achievable individual description rate can be derived by choosing a specific set of encoding auxiliary random variables, and more specifically, we shall choose the following set of random variables. Let

$$Y_{\alpha,k} = X + \sum_{i=\alpha}^{K-1} N_{i,k}, \quad \alpha = 1, 2, ..., K-1, \quad k = 1, 2, \ldots, K \tag{73}$$

where $N_{i,k}$ are mutually independent zero-mean Gaussian random variables, which are also independent of $X$. Their variances are denoted as $\sigma^2_{i,k}$, and they satisfy $\sigma^2_{i,k} = \sigma^2_{i,k'}$ for any $k, k' \in I_K$; we thus denote $\sigma^2_{i,k}$ as $\sigma^2_i$. For convenience, we shall denote $\sum_{i=\alpha}^{K-1} N_{i,k}$ as $Z_{\alpha,k}$. For the last layer, i.e., $\alpha = K$, we use

$$Y_K = X - \mathbb{E}(X|Y_{\alpha,k}, \alpha \in I_{K-1}, k \in I_K) + N_K, \tag{74}$$

where $N_K$ is a zero-mean Gaussian random variable independent of everything else, with variance $\sigma^2_K$. Clearly, $X - \mathbb{E}(X|Y_{\alpha,k}, \alpha \in I_{K-1}, k \in I_K)$ is the innovation of $X$ given all the lower-level random variables. It remains to specify the variances of $\{\{N_{\alpha,k}, \alpha \in I_{K-1}, k \in I_\alpha\}, N_K\}$, which is in fact not trivial as we shall discuss next. Notice that for all the layers except that $\alpha = K$-th layer, $Y_{\alpha-1,j} \leftrightarrow Y_{\alpha,j} \leftrightarrow X$ is a Markov string, thus the lower layers are useless when higher layers are decoded. To see that this choice of encoding auxiliary random variables does not satisfy the original symmetric distribution requirement, consider the joint distribution of $(Y_{1,1}, Y_{2,1})$ and that of $(Y_{1,1}, Y_{2,2})$. Given $X$, the first pair of random variables are dependent, while the second pair of random variables are independent; this clearly violates the original symmetric distribution requirement.

The key difficulty we face is now the following: when descriptions in the set $G_{\boldsymbol{v}}$, where $|\boldsymbol{v}| = \alpha - 1 \leq K - 1$, are received, the decoding function can reconstruct the source using the random variable $\{Y_{\alpha-1,i}, i \in G_{\boldsymbol{v}}\}$; note that from (73), it is clear that since $Y_{\alpha-2,k} = Y_{\alpha-1,k} + N_{\alpha-2,k}$, using only $\{Y_{\alpha-1,i}, i \in G_{\boldsymbol{v}}\}$ to reconstruct the source does not lose optimality, i.e., the lower layer random variables are useless given the higher layers. If one more description, say the $j$-th description, is further received, the decoding function now can utilize the random variable associated with this description $Y_{\alpha-1,j}$. Thus even if the $\alpha$-th layer random variables $\{Y_{\alpha,i}, i \in G_{\boldsymbol{v}} \cup \{j\}\}$ do not provide additional information beyond the lower layer random variables $\{Y_{\alpha-1,i}, i \in G_{\boldsymbol{v}} \cup \{j\}\}$, the decoder is still able to improve the reconstruction over the original decoding function with descriptions in $G_{\boldsymbol{v}}$. This is in fact a key observation in [5] that improves the system performance over the simple SR-ULP scheme. This observation implies that for certain distortion vector $\boldsymbol{D}$, it is not possible to satisfy all the constraints with equalities with the PPR multilayer scheme because some constraints are too loose, and thus the distortion region has some degenerate regimes. The enhanced distortion vector given in Section 4 is thus introduced to eliminate this effect. This enhanced distortion vector $\boldsymbol{D}^*$ serves a similar role as the enhanced channel in [19], where the MIMO Gaussian broadcast channel capacity is established.

The enhanced distortion vector $\boldsymbol{D}^*$ has the following three important properties:

- Enhancement: $\boldsymbol{D}^*$ enhanced the distortion vector $\boldsymbol{D}$, i.e., $D_i^* \leq D_i$, $i = 1, 2, ...K$.
- Monotonicity: $\boldsymbol{D}^* = (D_1^*, D_2^*, ..., D_K^*)$ is a monotonically decreasing sequence, thus a valid distortion vector.
- $\Phi$-monotonicity: it satisfies the condition

$$\Phi_\alpha(D_\alpha^*) \leq \Phi_{\alpha-1}(D_{\alpha-1}^*), \quad \alpha = 2, 3, ..., K. \tag{75}$$

These properties are straightforward to check by the construction of $D_k^*$.

The $\Phi$-monotonicity property is exactly the condition being checked in the definition of the enhanced distortion vector, with $D_\alpha^*$ replacing $D_\alpha$. Thus the definition of the enhanced distortion vector effectively constructs a new distortion vector in a sequential manner, if the original distortion vector does not satisfy the $\Phi$-monotonicity property. The desired $\Phi$-monotonicity property removes the degenerate regimes and the corresponding difficulty previously discussed. To see this, consider the following two cases: (1) when descriptions in $G_{\boldsymbol{u}}$ are received, where $|\boldsymbol{u}| = \alpha$; (2) when descriptions in $G_{\boldsymbol{v}}$ are received, where $\boldsymbol{v} = \alpha - 1$ and $G_{\boldsymbol{v}} \subseteq G_{\boldsymbol{u}}$. For the latter, using the given Gaussian auxiliary random variables $\{Y_{\alpha-1,i}, i \in G_{\boldsymbol{v}}\}$, linear estimation induces a distortion

$$D'_{\alpha-1} = \frac{\sum_{i=\alpha-1}^{K-1} \sigma_i^2}{\sum_{i=\alpha-1}^{K-1} \sigma_i^2 + \alpha - 1}. \tag{76}$$

Similarly, using the random variables $\{Y_{\alpha,i}, i \in G_u\}$, linear estimation induces

$$D'_\alpha = \frac{\sum_{i=\alpha}^{K-1} \sigma_i^2}{\sum_{i=\alpha}^{K-1} \sigma_i^2 + \alpha}. \tag{77}$$

In the case that each individual encoding auxiliary random variable $Y_{\alpha,j}$ does refine over $Y_{\alpha-1,j}$, i.e., there is no explicit information embedded in the $\alpha$-th layer, we have $\sigma_{\alpha-1}^2 = 0$, i.e., $D'_{\alpha-1}$ is given by

$$D'_{\alpha-1} = \frac{\sum_{i=\alpha}^{K-1} \sigma_i^2}{\sum_{i=\alpha}^{K-1} \sigma_i^2 + \alpha - 1}. \tag{78}$$

Now suppose the distortion constraint at the $(\alpha-1)$-th level is given by $D_{\alpha-1}=D'_{\alpha-1}$ as in (78), then the degenerate case previously discussed indeed occurs if the distortion constraint at the $\alpha$-th level is given such that $D_\alpha > D'_\alpha$. Through elementary algebra, it is clear that this is equivalent to the condition

$$\frac{\alpha D_\alpha}{1-D_\alpha} > \frac{(\alpha-1)D_{\alpha-1}}{1-D_{\alpha-1}}, \tag{79}$$

which is exactly the negation of (75), with $(D_{\alpha-1}, D_\alpha)$ replacing $(D^*_{\alpha-1}, D^*_\alpha)$.

Thus if the condition (75) does not hold for the given distortion constraints $D_{\alpha-1}$ and $D_\alpha$, our choice of Gaussian encoding auxiliary random variables will not be able to achieve the given $(D_{\alpha-1}, D_\alpha)$ simultaneously with equality, but can naturally achieve strictly better distortions with equality. For the enhanced distortion vector $(D^*_1, D^*_2, ..., D^*_K)$, which indeed satisfies the condition (75), the distortion constraints can always be satisfied with equality in this achievability scheme, by choosing the appropiate variances $(\sigma^2_1, \sigma^2_2, ..., \sigma^2_K)$. Conversely, given an enhanced distortion vector $\boldsymbol{D}^*$, the variances of the auxiliary random variables $\{\{N_{\alpha,k}, \alpha \in I_{K-1}, k \in I_K\}, N_K\}$ are uniquely determined. More precisely, the variances for $\sigma^2_\alpha$, $\alpha = 1, 2, ..., K-1$ are determined by

$$\sum_{i=\alpha}^{K-1} \sigma^2_i = \Phi_\alpha(D^*_\alpha), \tag{80}$$

which always give a set of valid choices of the variances. Thus from here on, in the PPR multilayer coding scheme, the Gaussian auxiliary random variables will be assumed to have the variances thus determined.

With the enhanced distortion vector $\boldsymbol{D}^*$ properly defined, we have the following corollary, the proof of which is given in Appendix 9.

**Corollary 5.2** *For the Gaussian source,*

$$R(\boldsymbol{D}) \le R(\boldsymbol{D}^*) \le \tilde{R}(\boldsymbol{D}^*). \tag{81}$$

The first inequality is clearly true because $\boldsymbol{D}^*$ enhances $\boldsymbol{D}$.

## 5.3 Lower bounding the sum rate

Next we generalize the lower bounding derivation given in Section 3 for $K=3$ to the case of general $K$. The generalization is notationally involved, and the result is summarized in the following theorem.

**Theorem 5.4** *For the Gaussian source, the sum rate under the $K$-description symmetric distortion satisfies*

$$\sum_{i=1}^{K} R_i \ge \frac{K}{2}\sum_{\alpha=1}^{K}\frac{1}{\alpha}\log\frac{(1+d_\alpha)(D_\alpha+d_{\alpha-1})}{(1+d_{\alpha-1})(D_\alpha+d_\alpha)}, \tag{82}$$

*where $d_1 \ge d_2 \ge ... \ge d_{K-1} > 0$ are arbitrary non-negative values, $d_0 \triangleq \infty$ and $d_K \triangleq 0$.*

**Proof 2** *The bounding technique extends the method used in [2, 8, 9], however with the new ingredient that we expand the probability space with more than one additional random variables, and then utilize the special* *structure in the expanded probability space to bound the sum rate. We have the following chain of inequalities*

$$
\begin{aligned}
n\sum_{i=1}^{K}(R_i+\epsilon) &\geq \sum_{i=1}^{K} H(S_i) - H(S_i, i\in I_K|X^n)\\
&\overset{(a)}{=} \sum_{\alpha=1}^{K-1}\left[\frac{K}{\alpha\binom{K}{\alpha}}\sum_{G_{\boldsymbol{v}}:|\boldsymbol{v}|=\alpha} H(S_i, i\in G_{\boldsymbol{v}}) - \frac{K}{(\alpha+1)\binom{K}{\alpha+1}}\sum_{G_{\boldsymbol{v}}:|\boldsymbol{v}|=\alpha+1} H(S_i, i\in G_{\boldsymbol{v}})\right]\\
&\qquad\qquad + H(S_i, i\in I_K) - H(S_i, i\in I_K|X^n)\\
&\overset{(b)}{\geq} I(S_i, i\in I_K; X^n)\\
&\quad + \sum_{\alpha=1}^{K-1}\left[\frac{K}{\alpha\binom{K}{\alpha}}\sum_{G_{\boldsymbol{v}}:|\boldsymbol{v}|=\alpha} H(S_i, i\in G_{\boldsymbol{v}}) - \frac{K}{(\alpha+1)\binom{K}{\alpha+1}}\sum_{G_{\boldsymbol{v}}:|\boldsymbol{v}|=\alpha+1} H(S_i, i\in G_{\boldsymbol{v}})\right]\\
&\quad - \sum_{\alpha=1}^{K-1}\left[\frac{K}{\alpha\binom{K}{\alpha}}\sum_{G_{\boldsymbol{v}}:|\boldsymbol{v}|=\alpha} H(S_i, i\in G_{\boldsymbol{v}}|Y_\alpha^n) - \frac{K}{(\alpha+1)\binom{K}{\alpha+1}}\sum_{G_{\boldsymbol{v}}:|\boldsymbol{v}|=\alpha+1} H(S_i, i\in G_{\boldsymbol{v}}|Y_\alpha^n)\right]\\
&= I(S_i, i\in I_K; X^n)\\
&\quad + \sum_{\alpha=1}^{K-1}\left[\frac{K}{\alpha\binom{K}{\alpha}}\sum_{G_{\boldsymbol{v}}:|\boldsymbol{v}|=\alpha} I(S_i, i\in G_{\boldsymbol{v}}; Y_\alpha^n) - \frac{K}{(\alpha+1)\binom{K}{\alpha+1}}\sum_{G_{\boldsymbol{v}}:|\boldsymbol{v}|=\alpha+1} I(S_i, i\in G_{\boldsymbol{v}}; Y_\alpha^n)\right]\\
&= \sum_{i=1}^{K} I(S_i; Y_1^n) + \sum_{\alpha=2}^{K-1}\frac{K}{\alpha\binom{K}{\alpha}}\sum_{G_{\boldsymbol{v}}:|\boldsymbol{v}|=\alpha}\left[I(S_i, i\in G_{\boldsymbol{v}}; Y_\alpha^n) - I(S_i, i\in G_{\boldsymbol{v}}; Y_{\alpha-1}^n)\right]\\
&\qquad\qquad + \left[I(S_i, i\in I_K; X^n) - I(S_i, i\in I_K; Y_{K-1}^n)\right].
\end{aligned}
\tag{83}
$$

*where (a) is by adding and subtracting the same terms where the positive term in the bracket chases the negative one; (b) is true because the subtracted bracket is nonnegative due to conditional version of Han's inequality [22];* $\{Y_\alpha, \alpha\in I_K\}$ *are defined in (66), though here we are not using them to construct codes. For convenience we denote* $d_\alpha = \sum_{j=\alpha}^{K}\sigma_j^2$*. Now we can apply Lemma 3.1 on (83) to get the desired result by noticing*

$$
\log\frac{D_1+d_0}{1+d_0} = \log\frac{D_1+\infty}{1+\infty} = 0, \tag{84}
$$

*with the convention* $\log\frac{\infty}{\infty}=0$.

Note that the lower bound in Theorem 5.4 is in fact a set of lower bounds, parametrized by $d_1 \geq d_2 \geq \ldots \geq d_{K-1} > 0$. We may optimize it to find the tightest lower bound, however, an explicit optimization is not only difficult, but also fails to offer much insight due to the lack of matching achievability result. Instead we shall choose a specific set of values to get a (sub-optimal) bound, resulting in the following corollary.

**Corollary 5.3** *For the Gaussian source, the SID-RD function under symmetric distortion constraints $\boldsymbol{D}$ satisfies*

$$\begin{aligned} R(\boldsymbol{D}) &\geq \frac{1}{2}\sum_{\alpha=1}^{K}\frac{1}{\alpha}\log\frac{D^*_{\alpha-1}}{D^*_{\alpha}} - \frac{1}{2}\sum_{\alpha=2}^{K}\frac{1}{\alpha-1}\log(\alpha - D^*_{\alpha-1}) + \frac{1}{2}\sum_{\alpha=2}^{K}\frac{1}{\alpha}\log(\alpha-1) \\ &\geq \frac{1}{2}\sum_{\alpha=1}^{K}\frac{1}{\alpha}\log\frac{D^*_{\alpha-1}}{D^*_{\alpha}} - \frac{1}{2}\sum_{\alpha=2}^{K}\frac{1}{\alpha-1}\log\alpha + \frac{1}{2}\sum_{\alpha=2}^{K}\frac{1}{\alpha}\log(\alpha-1), \end{aligned} \tag{85}$$

*where $\boldsymbol{D}^*$ is the enhanced distortion vector of $\boldsymbol{D}$.*

This corollary is proved in Appendix 10. It is worth noting that the left hand side of (85) is regarding the SID-RD function of distortion vector $\boldsymbol{D}$, and the right hand sides of (85) are only related to the enhanced distortion vector. Indeed the enhanced distortion vector is given in such a way that it does not change the lower bound under the chosen value of $(d_1, d_2, ..., d_{K-1})$.

## 5.4 Bounding the gap between lower and upper bounds

Now it is rather straightforward to prove Theorem 4.1. Since $\boldsymbol{D}^*$ enhances $\boldsymbol{D}$, we have by Corollary 5.1 and Corollary 5.2 that

$$R(\boldsymbol{D}) \leq R(\boldsymbol{D}^*) \leq \tilde{R}(\boldsymbol{D}^*) \leq \hat{R}(\boldsymbol{D}^*). \tag{86}$$

Now combining (86) with Theorem 5.4 and Corollary 5.3 gives Theorem 4.1.

Theorem 4.1 provides one possible approximation for the SID-RD function with universal constant bit bound. Various improvements can be made, for example, better choice of $(d_1, d_2, ..., d_{K-1})$ and better choice of random variables in the PPR multilayer coding scheme. Moreover, when proving Corollary 5.3, we have omitted many terms, which may make the bound looser. In fact, for the case with only two level distortion constraints, the outer bound in Theorem 5.4 reduces correctly to the one given in [8] and [9]. It was shown in [2], [8] and [9] that for certain cases this bound is indeed tight, which however requires optimization to find the optimal bound. We will not pursue such refinements here, but leave them to interested readers.

In order to prove Corollary 4.1, notice that this case implies we can choose $D_1 = D_2 = ... = D_{K-k} = 1$, and furthermore we can set $d_{K-k} = \infty$. Thus the lower bound $\underline{R}(\boldsymbol{D}, \boldsymbol{d})$ implies that

$$R(\boldsymbol{D}) \geq \frac{1}{2}\sum_{\alpha=K-k+1}^{K}\frac{1}{\alpha}\log\frac{(1+d_\alpha)(D_\alpha + d_{\alpha-1})}{(1+d_{\alpha-1})(D_\alpha + d_\alpha)}. \tag{87}$$

Apply the procedure of computing the enhanced distortion vector on $(D_{K-k+1}, D_{K-k+2}, .., D_K)$ only, and denote the output as $(D^*_{K-k+1}, D^*_{K-k+2}, ..., D^*_K)$. We then follow the proof of Corollary 5.3 and arrive at

$$\begin{aligned} R(\boldsymbol{D}) &\geq \frac{1}{2}\sum_{\alpha=K-k+1}^{K}\frac{1}{\alpha}\log\frac{D^*_{\alpha-1}}{D^*_{\alpha}} - \frac{1}{2}\sum_{\alpha=K-k+2}^{K}\frac{1}{\alpha-1}\log(\alpha - D^*_{\alpha-1}) + \frac{1}{2}\sum_{\alpha=K-k+2}^{K}\frac{1}{\alpha}\log(\alpha-1) \\ &\geq \frac{1}{2}\sum_{\alpha=K-k+1}^{K}\frac{1}{\alpha}\log\frac{D^*_{\alpha-1}}{D^*_{\alpha}} - \frac{1}{2}\sum_{\alpha=K-k+2}^{K}\frac{1}{\alpha-1}\log\alpha + \frac{1}{2}\sum_{\alpha=K-k+2}^{K}\frac{1}{\alpha}\log(\alpha-1). \end{aligned} \tag{88}$$

It is clear that $\tilde{H}_\alpha = 0$ for $\alpha = 1, 2, ..., K-k$. Thus we have proved the bound for the differences between the upper and lower bounds as given in Corollary 4.1.

### 5.5 Extension to general sources

In this subsection we generalize the result for the Gaussian source to other sources under the MSE distortion measure, and show similar but looser bounds hold for the symmetric individual description rate under the quadratic distortion measure. We derive the result using the SR-ULP scheme, but not the PPR multilayer scheme, which appears difficult to analyze for general sources. Interestingly, for $K = 2$ and the symmetric distortion constraints, the sum rate gap between the upper bound derived using the SR-ULP scheme and the R-D function is upper-bounded by $1.5$ bits, which is the same value as that derived in [23] for the two description case; nevertheless our result is a stronger, since in [23] the achievable scheme is more involved than the SR-ULP scheme yet the bounding constant is the same.

Some additional definitions are necessary. For a general source $X$ with finite differential entropy, zero mean and unit variance, define the following quantity,

$$\hat{R}'(\boldsymbol{D}) = \sum_{\alpha=1}^{K} \frac{1}{\alpha} I(X; Y_\alpha | Y_1, Y_2, \ldots, Y_{\alpha-1}) \tag{89}$$

where random variable $Y_\alpha$, $\alpha = 1, 2, ..., K$ are defined as in (66) and (67).

The following theorem is the main result of this section.

**Theorem 5.5** *For any general source $X$ with unit variance under the MSE distortion measure, we have*

$$\hat{R}'(\boldsymbol{D}) \geq R(\boldsymbol{D}), \tag{90}$$

*moreover,*

$$\hat{R}'(\boldsymbol{D}) - R(\boldsymbol{D}) \leq \sum_{\alpha=1}^{K} \frac{1}{2\alpha}. \tag{91}$$

This theorem essentially states the the SR-ULP scheme with the additive Gaussian codebook operates within $\sum_{\alpha=1}^{K} (2\alpha)^{-1}$ of the optimal coding scheme, in terms of individual description rate, for any source with unit variance. The first statement in the theorem is trivial by applying Theorem 5.1, and the second statement is proved in Appendix 11.

Unlike Theorem 4.1, there is no explicit lower bound on the SID-RD function. Indeed, in the proof of Theorem 5.5, the outer bound is never explicitly written to have an single letter form or an analytical form that can be computed directly. The key proof idea is to construct the lower and upper bound in appropriate forms such that certain terms are the same, and then cancel these terms to bound the remaining terms.

## 6 Rate-distortion region approximation

In this section, we develop the results further to provide an approximate characterization for the MD rate-distortion region. The main difficulties are as follows. Firstly, the PPR multilayer scheme was originally designed for the symmetric rate only instead for an achievable rate region, and thus certain generalization has to be introduced to "inflate" it to a rate region. We apply the $\alpha$-resolution method to assert that the achievability of the corner points of a region which matches the polytopic template of the MLD rate region, and therefore by a time-sharing argument provide an achievable region. The second difficulty is in generalizing the sum-rate lower bounding technique to other rate combinations. The terms in deriving the lower bound are well-structured for the sum rate case, however for general rate combinations the terms lack such structure. Unlike the case $K = 3$, there is no explicit method to enumerate the appropriate rate combinations, i.e., the bounding faces of the rate regions. To overcome this difficulty, we combine the $\alpha$-resolution method with the sum rate lower bounding technique to provide the outer bound, or rather the lower bound for the bounding planes of the rate region.

## 6.1 Achievable rate-distortion region by the SR-MLD scheme

Parallel to the SID-RD case, we give a general definition of the rate region not necessary using a Gaussian codebook, which is based on the SR-MLD coding scheme illustrated in Fig. 3. Let $\hat{\mathcal{R}}(\boldsymbol{Y})$ be the set of non-negative rate vectors $(R_1, R_2, ..., R_K)$, such that

$$R_i \geq \sum_{\alpha=1}^{K} r_i^{\alpha}, \quad 1 \leq i \leq K, \tag{92}$$

for some $r_i^{\alpha} \geq 0$, $1 \leq \alpha \leq K$, satisfying

$$\sum_{i \in G_{\boldsymbol{v}}} r_i^{|\boldsymbol{v}|} \geq I(X; Y_{|\boldsymbol{v}|} | Y_1, Y_2, \ldots, Y_{|\boldsymbol{v}|-1}), \quad \forall \boldsymbol{v} \in \Omega_K. \tag{93}$$

We have slightly abused the notation in the above definitions by letting the argument of $\hat{\mathcal{R}}(\cdot)$ be a fixed set of random variables rather than a set of distortion constraints; this however does not cause much confusion due to their apparent difference.

**Theorem 6.1** *We have*

$$\textbf{\textit{conv}}(\hat{\mathcal{R}}(\boldsymbol{Y})) \subseteq \mathcal{R}(\boldsymbol{D}), \tag{94}$$

*where* **conv**$(\cdot)$ *is the convex hull operator, and it is taken over the set of auxiliary random variables* $\boldsymbol{Y} = (Y_1, Y_2, ..., Y_K)$ *in some alphabets* $\mathcal{Y}_1 \times \mathcal{Y}_2 \times ... \times \mathcal{Y}_K$*, which are jointly distributed with* $X$*, such that there exist deterministic functions* $g_\alpha : \mathcal{Y}_1 \times \mathcal{Y}_2 \times ... \times \mathcal{Y}_\alpha \to \mathcal{X}$ *to satisfy*

$$\mathbb{E}d(X, g_\alpha(Y_1, Y_2, ..., Y_\alpha)) \leq D_\alpha, \quad \alpha = 1, 2, ..., K. \tag{95}$$

By choosing the auxiliary random variables $Y_\alpha$, $\alpha = 1, 2, ..., K$ as specified by (66) and (67), it is clear that $\hat{\mathcal{R}}(\boldsymbol{D})$ is a (proper) subset of **conv**$(\hat{\mathcal{R}}(\boldsymbol{Y}))$, and thus an achievable region. Note that the region **conv**$(\hat{\mathcal{R}}(\boldsymbol{Y}))$ may be a general convex region with curvy boundary, thus not a polytope. However $\hat{\mathcal{R}}(\boldsymbol{D})$ is a subset of this set by specializing it to a particular distribution, resulting in a polytope[4]. Interestingly it is not a contra-polymatroid as often encountered in multiuser information theory. A contra-polymatroid is usually defined as a mapping from subsets (of the rate indices) to a non-negative real number. However, here even for the three description case, there are four mappings associated with the set of all three rates, one bounding $R_1 + R_2 + R_3$, and the other three on the form of $2R_i + R_j + R_k$, thus does not result in a valid mapping. As such, the theory characterizing the vertex points of a contra-polymatroid does not offer simplification in the MD problem. The $\alpha$-resolution method invented in [16] is one approach to address this difficulty.

[4]It is the projection of the polytope $(R_1, R_2, ..., R_K, \{r_i^{\alpha}, \alpha \in I_K, i \in I_K\})$ on the first $K$ components.

## 6.2 Achievable rate-distortion region by the PPR multilayer scheme

In this subsection, we first briefly describe the PPR multilayer coding scheme, and then discuss the difficulties in generalizing this achievable symmetric rate result to an achievable rate region. To overcome these difficulties, we combine the $\alpha$-resolution method with an additional coding step to provide such an rate region.

The PPR multilayer coding scheme can be described roughly as follows. At layer $\alpha$, $\alpha \in I_{K-1}$ and for any description $k \in I_K$, codebooks of size $2^{nR'_{\alpha,k}}$ are generated using the marginal distribution of $Y_{\alpha,k}$. The rate $R'_{\alpha,k}$ should be sufficiently large such that for any source codeword, with high probability there exist codewords in the codebook $(\alpha, k)$, $\alpha \in I_{K-1}$ and $k \in I_K$ that are jointly typical with it. This can be done if we choose

$$R'_{\alpha,k} > h(Y_{\alpha,1}) - \frac{1}{K} h(Y_{\alpha,k}, k \in I_K | X, \{Y_{j,k}, j \in I_{\alpha-1}, k \in I_K\}). \tag{96}$$

Though there is no requirement that $R'_{\alpha,k} = R'_{\alpha,k'}$ for any $k \neq k'$, we intentionally make them equal to simplify the resulting achievable region; i.e., we choose

$$R'_{\alpha,k} = h(Y_{\alpha,1}) - \frac{1}{K} h(Y_{\alpha,k}, k \in I_K | X, \{Y_{j,k}, j \in I_{\alpha-1}, k \in I_K\}) + \delta, \tag{97}$$

for an arbitrarily small but positive $\delta$. Next codewords in a codebook are randomly and independently assigned into a total of $2^{nR_{\alpha,k}}$ bins, $\alpha \in I_{K-1}$ and $k \in I_K$. At the decoder, with any $k^*$ descriptions such that $k^* \in I_{K-1}$, the first $k^*$ layers are decoded. More precisely, the decoder receives descriptions in $G_{\boldsymbol{v}}$, such that $|\boldsymbol{v}| = k^*$; if there exists a unique set of codewords $\{y^n_{\alpha,j}, \alpha \in I_{k^*}, j \in G_{\boldsymbol{v}}\}$, in the corresponding bins that are jointly typical, then the decoder reconstructs using the single-letter decoding function $g_{\boldsymbol{v}}(\cdot)$; otherwise a decoding failure occurs. To succeed with high probability for any $k^* \in I_{K-1}$, the rates $R_{\alpha,k}$, $\alpha \in I_{K-1}$ and $k \in I_K$, only need to satisfy

$$0 \leq R_{\alpha,j} \leq R'_{\alpha,j}, \quad \alpha \in I_{K-1}, \quad j \in I_K. \tag{98}$$

and

$$\sum_{j \in G_{\boldsymbol{v}}} (R'_{\alpha,j} - R_{\alpha,j}) < \alpha h(Y_{\alpha,1}) - h(Y_{\alpha,i}, i \in I_\alpha | Y_{k,j}, k \in I_{\alpha-1}, j \in I_\alpha), \tag{99}$$

for all $\boldsymbol{v} \in \Omega_K$ such that $|\boldsymbol{v}| = \alpha$, and for all $\alpha \in I_{K-1}$. Rewriting (99), we have

$$\begin{aligned}
\sum_{j \in G_{\boldsymbol{v}}} R_{\alpha,j} \quad \geq & \quad \sum_{j \in G_{\boldsymbol{v}}} R'_{\alpha,j} - \alpha h(Y_{\alpha,1}) + h(Y_{\alpha,i}, i \in I_\alpha | Y_{k,j}, k \in I_{\alpha-1}, j \in I_\alpha) \\
= & \quad \alpha h(Y_{\alpha,1}) - \frac{\alpha}{K} h(Y_{\alpha,k}, k \in I_K | X, \{Y_{j,k}, j \in I_{\alpha-1}, k \in I_K\}) + \alpha\delta \\
& \quad -\alpha h(Y_{\alpha,1}) + h(Y_{\alpha,i}, i \in I_\alpha | Y_{k,j}, k \in I_{\alpha-1}, j \in I_\alpha) \\
= & \quad h(Y_{\alpha,i}, i \in I_\alpha | Y_{k,j}, k \in I_{\alpha-1}, j \in I_\alpha) \\
& \quad -\frac{\alpha}{K} h(Y_{\alpha,k}, k \in I_K | X, \{Y_{j,k}, j \in I_{\alpha-1}, k \in I_K\}) + \alpha\delta.
\end{aligned} \tag{100}$$

The last layer codebook is generated using the more conventional method, i.e., the conditional codebook, and the following condition is sufficient

$$\sum_{k=1}^{K} R_{K,k} > I(X; Y_K | Y_{\alpha,k}, \alpha \in I_{K-1}, k \in I_K). \tag{101}$$

By collecting the constraints on non-negative rates $R_{\alpha,j}$ in (98), (100) and (101), and defining $R_k = \sum_{\alpha=1}^{K} R_{\alpha,k}$, we can already form an achievable region. However, the upper bound in (98) introduces additional difficulty when comparing to the outer bound derived in the next section, and thus it will be desirable to remove this condition. In other words, with these constraints taken into consideration, it is not clear whether the resulting region matches the polytopic template of the MLD coding rate region. Next we define a similar region, and prove this region is indeed achievable and can be written in a form with the same structure as the desired template. In [25], we gave a different scheme by using orthogonal binning, however we believe the scheme given below is more straightforward. We first introduce a few more notations.

For a fixed set of (generalized symmetric) auxiliary random variables $\{\{Y_{\alpha,k}, \alpha \in I_{K-1}, k \in I_K\}, Y_K\}$, define the following quantities for $\alpha \in I_{K-1}$,

$$\tilde{H}_\alpha(\boldsymbol{Y}) = h(Y_{\alpha,i}, i \in I_\alpha | Y_{k,j}, k \in I_{\alpha-1}, j \in I_\alpha) - \frac{\alpha}{K} h(Y_{\alpha,k}, k \in I_K | X, \{Y_{j,k}, j \in I_{\alpha-1}, k \in I_K\}), \quad (102)$$

and

$$\tilde{H}_K(\boldsymbol{Y}) = I(X; Y_K | Y_{\alpha,k}, \alpha \in I_{K-1}, k \in I_K). \quad (103)$$

Let $\tilde{\mathcal{R}}(\boldsymbol{Y})$ be the set of non-negative rate vectors $(R_1, R_2, ..., R_K)$, such that

$$R_i \geq \sum_{\alpha=1}^{K} r_i^\alpha, \quad 1 \leq i \leq K, \quad (104)$$

for some $r_i^\alpha \geq 0$, $1 \leq \alpha \leq K$, satisfying

$$\sum_{i \in G_{\boldsymbol{v}}} r_i^{|\boldsymbol{v}|} \geq \tilde{H}_{|\boldsymbol{v}|}(\boldsymbol{Y}), \quad \boldsymbol{v} \in \Omega_K. \quad (105)$$

Note here in fact the set of auxiliary random variables has more than $K$ components, however we still write it as $\boldsymbol{Y}$ for conciseness; we also slightly abuse the notation by letting $\tilde{H}_\alpha(\cdot)$ have either the enhanced distortion vector $\boldsymbol{D}^*$ or a set of random variables $\boldsymbol{Y}$ as the argument, which is indeed justified as we shall show that they are in fact the same by appropriate choice of the random variables $\boldsymbol{Y}$. The region $\tilde{\mathcal{R}}(\boldsymbol{Y})$ is the rate region satisfying (100), (101), and the lower bounds in (98), but not necessarily satisfying the upper bounds in (98). Thus for a fixed set of random variables $\{\{Y_{\alpha,k}, \alpha \in I_{K-1}, k \in I_K\}, Y_K\}$ and the specific choice of $R'_{\alpha,k}$, the achievable region directly implied by the PPR multilayer scheme, i.e., the one by collecting the constraints on non-negative rates $R_{\alpha,j}$ in (98), (100) and (101), is a subset of $\tilde{\mathcal{R}}(\boldsymbol{Y})$. We now state the following theorem.

**Theorem 6.2** *We have*

$$\textbf{\textit{conv}}(\tilde{\mathcal{R}}(\boldsymbol{Y})) \subseteq \mathcal{R}(\boldsymbol{D}), \quad (106)$$

*where the convex hull operator is taken over the set of generalized symmetric auxiliary random variables* $\{\{Y_{\alpha,k}, \alpha \in I_{K-1}, k \in I_K\}, Y_K\}$ *in the alphabets* $\mathcal{Y}_1^K \times \mathcal{Y}_2^K \times ... \times \mathcal{Y}_{K-1}^K \times \mathcal{Y}_K$*, which are jointly distributed with* $X$*, such that there exist deterministic functions* $g_\alpha : \mathcal{Y}_1^\alpha \times \mathcal{Y}_2^\alpha \times ... \times \mathcal{Y}_\alpha^\alpha \to \mathcal{X}$, $\alpha \in I_{K-1}$ *such that*

$$\mathbb{E}d(X, g_\alpha(\{Y_{i,k}, i \in I_\alpha, k \in I_\alpha\})) \leq D_\alpha, \quad \alpha = 1, 2, ..., K-1, \quad (107)$$

*and* $g_K : \mathcal{Y}_1^K \times \mathcal{Y}_2^K \times ... \times \mathcal{Y}_{K-1}^K \times \mathcal{Y}_K \to \mathcal{X}$*, such that*

$$\mathbb{E}d(X, g_K(\{Y_{i,k}, i \in I_\alpha, k \in I_\alpha\}, Y_K)) \leq D_K. \quad (108)$$

The proof of Theorem 6.2 relies on a result in [16], which is quoted below. Let "$\cdot$" denote the usual inner product in the Euclidean space. Let $\tilde{\mathcal{R}}^*(\boldsymbol{Y})$ be the set of all $\boldsymbol{R} \geq 0$ such that for all $\boldsymbol{A} \in \mathbb{R}_+^K$ but $\boldsymbol{A} \neq 0$

$$\boldsymbol{A} \cdot \boldsymbol{R} \geq \sum_{\alpha=1}^{K} f_\alpha(\boldsymbol{A}) \tilde{H}_\alpha(\boldsymbol{Y}), \tag{109}$$

where $f_\alpha(\boldsymbol{A})$ is defined in (16) of Section 2.C.

**Theorem 6.3 ([16] Theorem 2)**

$$\tilde{\mathcal{R}}(\boldsymbol{Y}) = \tilde{\mathcal{R}}^*(\boldsymbol{Y}). \tag{110}$$

*Remark:* In the definition of $\tilde{\mathcal{R}}^*(\boldsymbol{Y})$, the requirement that $\boldsymbol{R} \geq 0$ can be safely removed without loss of generality when $\tilde{H}_\alpha(\boldsymbol{Y}) \geq 0$. To see this, let $\boldsymbol{A} = (1, 0, ..., 0)$, then (109) reduces to $R_1 \geq \tilde{H}_1(\boldsymbol{Y})$ by applying Lemma 2.4.

In order to prove Theorem 6.2, consider the following. For a fixed set of (generalized symmetric) random variables $\{\{Y_{\alpha,k}, \alpha \in I_{K-1}, k \in I_K\}, Y_K\}$, since both $\mathcal{R}(\boldsymbol{D})$ and $\tilde{\mathcal{R}}(\boldsymbol{Y})$ are convex, they can be characterized by the bounding planes. As such if we can prove that for any $\boldsymbol{A} \in \mathbb{R}_+^K$ and $\boldsymbol{A} \neq 0$, the following inequality holds

$$\min_{\boldsymbol{R} \in \mathcal{R}(\boldsymbol{D})} \boldsymbol{A} \cdot \boldsymbol{R} \leq \min_{\boldsymbol{R} \in \tilde{\mathcal{R}}(\boldsymbol{Y})} \boldsymbol{A} \cdot \boldsymbol{R}, \tag{111}$$

then it follows that the region $\tilde{\mathcal{R}}(\boldsymbol{Y})$ is an achievable region. By Theorem 6.3, we have

$$\min_{\boldsymbol{R} \in \tilde{\mathcal{R}}(\boldsymbol{Y})} \boldsymbol{A} \cdot \boldsymbol{R} = \sum_{\alpha=1}^{K} f_\alpha(\boldsymbol{A}) \tilde{H}_\alpha(\boldsymbol{Y}). \tag{112}$$

Thus it suffices to prove that there always exists a rate vector in the achievable rate region that satisfy (112) with equality, i.e., there exists $\boldsymbol{R} \in \mathcal{R}(\boldsymbol{D})$ such that

$$\boldsymbol{A} \cdot \boldsymbol{R} = \sum_{\alpha=1}^{K} f_\alpha(\boldsymbol{A}) \tilde{H}_\alpha(\boldsymbol{Y}), \tag{113}$$

for any $\boldsymbol{A} \in \mathbb{R}_+^K$ and $\boldsymbol{A} \neq 0$. This would imply (111), which further implies the claimed result. We prove (113) and subsequently Theorem 6.2 in Appendix 12.

Notice that the region $\tilde{R}(\boldsymbol{D})$ is just $\tilde{R}(\boldsymbol{Y})$ with $\{\{Y_{\alpha,k}, \alpha \in I_{K-1}, k \in I_K\}, Y_K\}$ defined by (73) and (74), the variances of which are given by (80). Since for this specific choice of random variables, the values of $\tilde{H}_\alpha(\boldsymbol{Y}) = \tilde{H}_\alpha(\boldsymbol{D^*})$, $\alpha = 1, 2, ..., K$ are given in the proof of Corollary 5.2 in Appendix 9, the following corollary is now straightforward.

**Corollary 6.1** *Let $\boldsymbol{D^*}$ be the enhanced distortion vector of $\boldsymbol{D}$, then for the Gaussian source*

$$\tilde{\mathcal{R}}(\boldsymbol{D^*}) \subseteq \mathcal{R}(\boldsymbol{D^*}) \subseteq \mathcal{R}(\boldsymbol{D}). \tag{114}$$

## 6.3 Outer bounding the rate-distortion region

In this subsection, we provide a lower bound to the bounding plane of the Gaussian MD rate-distortion region.

**Theorem 6.4** *For the Gaussian source and any $\boldsymbol{A} \geq 0$,*

$$\sum_{\alpha=1}^{K} A_i R_i \geq \frac{1}{2} \sum_{\alpha=1}^{K} f_\alpha(\boldsymbol{A}) \log \frac{(1+d_\alpha)(D_\alpha + d_{\alpha-1})}{(1+d_{\alpha-1})(D_\alpha + d_\alpha)}. \tag{115}$$

*where the function $f_\alpha(\boldsymbol{A})$ is defined in (16), $d_1 \geq d_2 \geq ... \geq d_{K-1} > 0$ are arbitrary non-negative values, $d_0 \triangleq \infty$ and $d_K \triangleq 0$.*

**Proof 3** *Recall the result in Lemma 2.4, and consider the following inequalities,*

$$n \sum_{i=1}^{K} A_i (R_i + \epsilon) \geq \sum_{i=1}^{K} A_i H(S_i). \tag{116}$$

*Let $c_1, c_2, ..., c_K$ be a set of $\alpha$-resolution as defined in Theorem 2.1. Then we can write*

$$\begin{aligned}
\sum_{i=1}^{K} A_i H(S_i) &\overset{(a)}{=} \sum_{\alpha=1}^{K-1} \left[ \sum_{G_{\boldsymbol{v}}:|\boldsymbol{v}|=\alpha} c_\alpha(\boldsymbol{v}) H(S_i, i \in G_{\boldsymbol{v}}) - \sum_{G_{\boldsymbol{v}}:|\boldsymbol{v}|=\alpha+1} c_{\alpha+1}(\boldsymbol{v}) H(S_i, i \in G_{\boldsymbol{v}}) \right] \\
&\qquad + A_{\min} H(S_i, i \in I_K) - A_{\min} H(S_i, i \in I_K | X^n) \\
&\overset{(b)}{\geq} A_{\min} I(S_i, i \in I_K; X^n) \\
&\qquad + \sum_{\alpha=1}^{K-1} \left[ \sum_{G_{\boldsymbol{v}}:|\boldsymbol{v}|=\alpha} c_\alpha(\boldsymbol{v}) H(S_i, i \in G_{\boldsymbol{v}}) - \sum_{G_{\boldsymbol{v}}:|\boldsymbol{v}|=\alpha+1} c_{\alpha+1}(\boldsymbol{v}) H(S_i, i \in G_{\boldsymbol{v}}) \right] \\
&\qquad - \sum_{\alpha=1}^{K-1} \left[ \sum_{G_{\boldsymbol{v}}:|\boldsymbol{v}|=\alpha} c_\alpha(\boldsymbol{v}) H(S_i, i \in G_{\boldsymbol{v}} | Y_\alpha^n) - \sum_{G_{\boldsymbol{v}}:|\boldsymbol{v}|=\alpha+1} c_{\alpha+1}(\boldsymbol{v}) H(S_i, i \in G_{\boldsymbol{v}} | Y_\alpha^n) \right] \\
&= A_{\min} I(S_i, i \in I_K; X^n) \\
&\qquad + \sum_{\alpha=1}^{K-1} \left[ \sum_{G_{\boldsymbol{v}}:|\boldsymbol{v}|=\alpha} c_\alpha(\boldsymbol{v}) I(S_i, i \in G_{\boldsymbol{v}}; Y_\alpha^n) - \sum_{G_{\boldsymbol{v}}:|\boldsymbol{v}|=\alpha+1} c_{\alpha+1}(\boldsymbol{v}) I(S_i, i \in G_{\boldsymbol{v}}; Y_\alpha^n) \right] \\
&= \sum_{i=1}^{K} A_i I(S_i; Y_1^n) + \sum_{\alpha=2}^{K-1} \sum_{G_{\boldsymbol{v}}:|\boldsymbol{v}|=\alpha} c_\alpha(\boldsymbol{v}) \left[ I(S_i, i \in G_{\boldsymbol{v}}; Y_\alpha^n) - I(S_i, i \in G_{\boldsymbol{v}}; Y_{\alpha-1}^n) \right] \\
&\qquad + A_{\min} \left[ I(S_i, i \in I_K; X^n) - I(S_i, i \in I_K; Y_{K-1}^n) \right],
\end{aligned} \tag{117}$$

*where (a) is by adding and subtracting the same terms, and due to the fact that $S_i, i \in I_K$ are deterministic functions of $X^n$, and (b) is by a conditional version of the covering property of the given sequence of the optimal $\alpha$-resolutions as defined in (18). At this point, the expression is quite similar to (83), and we can apply Lemma 3.1 to complete the proof.*

Parallel to the sum rate case, we can specialize the lower bound by choosing the values of $d_1, d_2, ..., d_{K-1}$.

**Corollary 6.2** *For the Gaussian source, we have*

$$\sum_{i=1}^{K} A_i R_i \geq \frac{1}{2}\sum_{\alpha=1}^{K} f_\alpha(\boldsymbol{A}) \log \frac{D^*_{\alpha-1}}{D^*_\alpha} - \frac{1}{2}\sum_{\alpha=2}^{K} f_{\alpha-1}(\boldsymbol{A}) \log(\alpha - D^*_{\alpha-1}) + \frac{1}{2}\sum_{\alpha=2}^{K} f_\alpha(\boldsymbol{A}) \log(\alpha - 1) \tag{118}$$

$$\geq \frac{1}{2}\sum_{\alpha=1}^{K} f_\alpha(\boldsymbol{A}) \log \frac{D^*_{\alpha-1}}{D^*_\alpha} - \frac{1}{2}\sum_{\alpha=2}^{K} f_{\alpha-1}(\boldsymbol{A}) \log \alpha + \frac{1}{2}\sum_{\alpha=2}^{K} f_\alpha(\boldsymbol{A}) \log(\alpha - 1), \tag{119}$$

*where $\boldsymbol{D}^*$ is the enhanced distortion vector of $\boldsymbol{D}$.*

The proof is given Appendix 13, which is along a similar line as the proof of Corollary 5.3, with the additional application of Lemma 2.3 in one step.

Next we proceed to establish that the outer bound given above is indeed a polytope. More precisely, define $\mathcal{R}_L(\boldsymbol{D}^*)$ to be the set of $\boldsymbol{R} \in \mathbb{R}^K$, such that (118) holds for any $\boldsymbol{A} \in \mathbb{R}^K_+$ and $\boldsymbol{A} \neq 0$. Note that we do not require $R_i \geq 0$ in this set. The following corollary establishes a polytopic outer bound.

**Corollary 6.3** *Let $\boldsymbol{D}^*$ be the enhanced distortion vector of $\boldsymbol{D}$, then $\mathcal{R}_L(\boldsymbol{D}^*) \cap \mathbb{R}^K_+$ is a polytope such that $\mathcal{R}(\boldsymbol{D}) \subseteq \mathcal{R}_L(\boldsymbol{D}^*) \cap \mathbb{R}^K_+$.*

The proof of this corollary is given in Appendix 14. The key idea is the following: though we have an uncountable number of bounding planes to characterize $\mathcal{R}_L(\boldsymbol{D}^*)$, if there exists a set $\mathcal{S}_R \subseteq \mathcal{R}_L(\boldsymbol{D}^*)$ with finite number of elements, such that for each $\boldsymbol{A}$, inequality (118) can be satisfied with equality for some element in $\mathcal{S}_R$, then $\mathcal{R}_L(\boldsymbol{D}^*)$ is a polytope. The proof given in Appendix 14 proves the existence of such a finite set.

## 6.4 Bounding the gap between outer and inner bounds

Now we are ready to prove Theorem 4.2 and Theorem 4.3, which are presented below.

**Proof 4 (Proof of Theorem 4.2 and Theorem 4.3)** *Theorem 4.2 is implied by Theorem 6.1, Theorem 6.2 (or rather Corollary 6.1), the fact that $\boldsymbol{D}^*$ enhances $\boldsymbol{D}$, and the fact that for $\alpha = 2, 3, ..., K$*

$$\tilde{H}_\alpha(\boldsymbol{D}^*) = \frac{1}{2}\log\frac{(\alpha-1)D^*_{\alpha-1}}{D^*_\alpha(\alpha - D^*_{\alpha-1})} \leq \frac{1}{2}\log\frac{D^*_{\alpha-1}}{D^*_\alpha}, \tag{120}$$

*and $\tilde{H}_1(\boldsymbol{D}^*) = \frac{1}{2}\log\frac{1}{D^*_1}$.*

*The first inequality in (61) can be proved by (119) and the definition of $\hat{H}(\boldsymbol{D}^*)$, and invoking Theorem 4.2, Theorem 6.3, Theorem 6.4 and Corollary 6.2. To prove the first inequality in (62) of Theorem 4.3, we again* *combine Theorem 4.2, Theorem 6.3, Theorem 6.4, Corollary 6.2, and notice the following fact*

$$
\begin{aligned}
&\sum_{\alpha=1}^{K} f_\alpha(\boldsymbol{A})\tilde{H}_\alpha \\
&= \frac{1}{2} f_1(\boldsymbol{A}) \log \frac{1}{D_1^*} + \frac{1}{2}\sum_{\alpha=2}^{K-1} f_\alpha(\boldsymbol{A}) \log \frac{(\alpha-1)D_{\alpha-1}^*}{D_\alpha^*(\alpha - D_{\alpha-1}^*)} + \frac{1}{2} f_K(\boldsymbol{A}) \log \frac{(K-1)D_{K-1}^*}{D_K^*(K-D_{K-1}^*)} \\
&= \frac{1}{2}\sum_{\alpha=1}^{K} f_\alpha(\boldsymbol{A}) \log \frac{D_{\alpha-1}^*}{D_\alpha^*} - \frac{1}{2}\sum_{\alpha=2}^{K} f_\alpha(\boldsymbol{A}) \log(\alpha - D_{\alpha-1}^*) + \frac{1}{2}\sum_{\alpha=2}^{K} f_\alpha(\boldsymbol{A}) \log(\alpha-1).
\end{aligned} \tag{121}
$$

*The first inequality in (62) now follows from (118) and the definition of $\tilde{H}(\boldsymbol{D}^*)$.*

*To prove the second inequality in (61), we write*

$$
\begin{aligned}
&\frac{1}{2}\sum_{\alpha=2}^{K} f_{\alpha-1}(\boldsymbol{A}) \log \alpha - \frac{1}{2}\sum_{\alpha=2}^{K} f_\alpha(\boldsymbol{A}) \log(\alpha-1) \\
&\quad = \frac{f_1(\boldsymbol{A})}{2}\log 2 + \frac{1}{2}\sum_{\alpha=2}^{K-1} f_\alpha(\boldsymbol{A})[\log(\alpha+1) - \log(\alpha-1)] - \frac{f_K(\boldsymbol{A})}{2}\log(K-1) \\
&\quad \overset{(a)}{\leq} \frac{A_{sum}}{2}\log 2 + \frac{1}{2}\sum_{\alpha=2}^{K-1} \frac{A_{sum}}{\alpha}[\log(\alpha+1) - \log(\alpha-1)] - \frac{A_{\min}}{2}\log(K-1) \\
&\quad \leq \frac{A_{sum}}{2}\sum_{\alpha=2}^{K} \frac{1}{\alpha-1}\log\alpha - \frac{A_{sum}}{2}\sum_{\alpha=2}^{K-1}\frac{1}{\alpha}\log(\alpha-1) - \frac{A_{\min}}{2}\log(K-1),
\end{aligned} \tag{122}
$$

*where in (a) we use Lemma 2.4. The second inequality in (62) can be proved similarly, and the details are omitted.*

## 6.5 Extension to general sources

Similar to the SID-RD approximation, we can extend the rate-distortion region approximation technique to general sources under the MSE distortion measure. It is clear that the definition of $\hat{\mathcal{R}}(\boldsymbol{Y})$ is not limited to the Gaussian source, and denote $\hat{\mathcal{R}}'(\boldsymbol{D})$ as $\hat{\mathcal{R}}(\boldsymbol{Y})$ with the random variables $\boldsymbol{Y}$ defined as (66) and (67). Define the following function,

$$\hat{R}'_{\boldsymbol{A}}(\boldsymbol{D}) \triangleq \min_{\boldsymbol{R}\in\hat{\mathcal{R}}'(\boldsymbol{D})} \boldsymbol{A}\cdot\boldsymbol{R} \tag{123}$$

We have the following theorem.

**Theorem 6.5** *For any general source $X$ with unit variance under the MSE distortion measure, we have*

$$\hat{\mathcal{R}}'(\boldsymbol{D}) \subseteq \mathcal{R}(\boldsymbol{D}), \tag{124}$$

*moreover, for any $\boldsymbol{A} \in \mathbb{R}_+^K$ and $\boldsymbol{A} \neq 0$*

$$\hat{R}'_{\boldsymbol{A}}(\boldsymbol{D}) - R_{\boldsymbol{A}}(\boldsymbol{D}) \leq \sum_{\alpha=1}^{K} \frac{f_\alpha(\boldsymbol{A})}{2}. \tag{125}$$

The proof follows closely the sum rate case proof for general sources, and we thus omit it here.

# 7 Conclusion

We provide approximate characterizations of the individual-description R-D function, as well as the achievable rate region, for the Gaussian MD problem under symmetric distortion constraints. This is done by combining two inter-connected parts: the derivation of a novel outer bound, and careful analysis of achievability schemes to generate inner bounds for easy comparison with the outer bound. The outer bound alone, or the inner bound alone, will not be able to provide this results, and particular care has to be taken in order to make them compatible. A result in a similar vein was recently obtained by Etkin *et al.* [26] for Gaussian interference channel.

The new lower bound is obtained by generalizing Ozarow's well-known technique, and expand the probability space of the original problem by more than one random variables with special structure among them. This technique appears to be promising, and we expect to see its application in other difficult multi-terminal information-theoretic communication problems.

The multi-level diversity coding problem, which can be understood as a lossless counterpart of the MD problem, shed tremendous light on the geometric structure of the MD rate-distortion region. We use the lossless MLD coding rate region as a polytopic template for both inner and outer bounds for the MD rate-distortion region. With the increasing complexity of a source coding problem being considered in information theory literature, we expect the complexity of its lossless counterpart to increase as well, and the difficulty of the corresponding lossless problem becomes an increasingly dominant component of the overall problem. In this context, our work can be understood as the first attempt to make explicit connection between the lossless source coding problem and its lossy counterpart. It is worth noting in multi-terminal channel coding problems, several well-known recent works can be understood as using deterministic models, for example, the network coding results in [27], and the deterministic wireless relay channel model in [28]. There exists a philosophical connection between the approach taken in this work and the "one-bit" approximation result for the Gaussian interference channel in [26], as well as the approximate capacity result for the Gaussian relay network [29]. In [29], an approximate characterization was motivated by the insight obtained in studying deterministic relay networks [28], which has an analogous role as the lossless multi-level diversity coding problem in our work. In both cases, the connection provides useful insight to the coding scheme and outer bounding proof technique. We expect in the near future connection between the lossless (deterministic) model and their lossy (non-deterministic) counterpart to be made on other information theoretic problems, and the approach of using the former as a guideline in treating the latter to be a fruitful path.

# 8 Proof of Lemma 3.1

**Proof 5** *Define $Z_a = N_a + N_b$ and $Z_b = N_b$. To prove the first statement, we consider the following chain of inequalities*

$$
\begin{aligned}
&I(S_i, i \in G_{\boldsymbol{v}}; Y_a^n) \\
&= nh(Y_a) - h(Y_a^n | S_i, i \in G_{\boldsymbol{v}}) \\
&= nh(Y_a) - h(X^n + Z_a^n | S_i, i \in G_{\boldsymbol{v}}) \\
&= nh(Y_a) - h(X^n + Z_a^n - \hat{X}_{\boldsymbol{v}}^n | S_i, i \in G_{\boldsymbol{v}}) \\
&\overset{(a)}{\geq} nh(Y_a) - h(X^n + Z_a^n - \hat{X}_{\boldsymbol{v}}^n) \\
&\overset{(b)}{\geq} nh(Y_a) - \sum_{i=1}^{n} h[X(i) + Z_a(i) - \hat{X}_{\boldsymbol{v}}(i)] \\
&\overset{(c)}{\geq} nh(Y_a) - \sum_{i=1}^{n} \frac{1}{2} \log \left\{ (2\pi e)\mathbb{E}[(X(i) + Z_a(i) - \hat{X}_{\boldsymbol{v}}(i))^2] \right\} \\
&= nh(Y_a) - \sum_{i=1}^{n} \frac{1}{2} \log \left[ (2\pi e)(\mathbb{E}d(X(i), \hat{X}_{\boldsymbol{v}}(i)) + d_a) \right],
\end{aligned}
$$

*where $\hat{X}_{\boldsymbol{v}}^n$ is the reconstruction with descriptions $S_i, i \in G_{\boldsymbol{v}}$, and its $i$-th position is denoted as $\hat{X}_{\boldsymbol{v}}(i)$. The inequality (a) is because conditioning reduces entropy, (b) is because of the chain rule for differential entropy and the fact that conditioning reduces entropy, and (c) is because Gaussian distribution maximizes the differential entropy for a given second moment. Since $\log(\cdot)$ is a concave function, we have*

$$
\sum_{i=1}^{n} \frac{1}{2} \log \left[ (2\pi e)(\mathbb{E}d(X(i), \hat{X}_{\boldsymbol{v}}(i)) + d_a) \right] \leq \frac{n}{2} \log \left( 2\pi e \mathbb{E}d(X^n, \hat{X}_{\boldsymbol{v}}^n) + d_a \right).
$$

*And it follows*

$$
\begin{aligned}
I(S_i, i \in G_{\boldsymbol{v}}; Y_a^n) &\geq nh(Y_a) - \frac{n}{2} \log \left( 2\pi e \mathbb{E}d(X^n, \hat{X}_{\boldsymbol{v}}^n) + d_a \right) \\
&\geq nh(Y_a) - \frac{n}{2} \log \left( (2\pi e)(D_{|\boldsymbol{v}|} + d_a) \right) \\
&= \frac{n}{2} \log \frac{1 + d_a}{D_{|\boldsymbol{v}|} + d_a},
\end{aligned}
$$

*which is the first claim.*

*To prove the second claim, we write the following*

$$
\begin{aligned}
&I(S_i, i \in G_{\boldsymbol{v}}; Y_b^n) - I(S_i, i \in G_{\boldsymbol{v}}; Y_a^n) \\
&= nh(Y_b) - nh(Y_a) + h(Y_a^n | S_i, i \in G_{\boldsymbol{v}}) - h(Y_b^n | S_i, i \in G_{\boldsymbol{v}}).
\end{aligned}
$$

*For the latter two terms, we have*

$$\begin{aligned}
&h(Y_a^n|S_i, i \in G_{\boldsymbol{v}}) - h(Y_b^n|S_i, i \in G_{\boldsymbol{v}}) \\
&\overset{(a)}{=} h(Y_a^n|S_i, i \in G_{\boldsymbol{v}}) - h(Y_b^n|N_a^n, \{S_i, i \in G_{\boldsymbol{v}}\}) \\
&\overset{(b)}{=} h(Y_a^n|S_i, i \in G_{\boldsymbol{v}}) - h(Y_a^n|N_a^n, \{S_i, i \in G_{\boldsymbol{v}}\}) \\
&= I(Y_a^n; N_a^n|S_i, i \in G_{\boldsymbol{v}}),
\end{aligned}$$

*where (a) is because $N_a^n$ is independent of $Y_b^n$ and $\{S_i, i \in G_{\boldsymbol{v}}\}$; (b) is by the definition of $Y_a$. Continuing along this line, we have*

$$\begin{aligned}
&I(Y_a^n; N_a^n|S_i, i \in G_{\boldsymbol{v}}) \\
&\overset{(a)}{=} h(N_a^n) - h(N_a^n|X^n + N_a^n + N_b^n, \{S_i, i \in G_{\boldsymbol{v}}\}) \\
&= h(N_a^n) - h(N_a^n|X^n + N_b^n + N_a^n, \hat{X}_{\boldsymbol{v}}^n, \{S_i, , i \in G_{\boldsymbol{v}}\}) \\
&\overset{(b)}{\geq} h(N_a^n) - h(N_a^n|X^n - \hat{X}_{\boldsymbol{v}}^n + N_a^n + N_b^n) \\
&\overset{(c)}{\geq} \sum_{i=1}^{n} \Big( h(N_a(i)) - h(N_a(i)|X(i) - \hat{X}_{\boldsymbol{v}}(i) + N_a(i) + N_b(i) \Big) \\
&= \sum_{i=1}^{n} I(N_a(i); X(i) - \hat{X}_{\boldsymbol{v}}(i) + N_b(i) + N_a(i)) \\
&\overset{(d)}{\geq} \sum_{i=1}^{n} \frac{1}{2} \log \frac{\mathbb{E}d(X(i), \hat{X}_{\boldsymbol{v}}(i)) + d_a}{\mathbb{E}d(X(i), \hat{X}_{\boldsymbol{v}}(i)) + d_b} \\
&\overset{(e)}{\geq} \frac{n}{2} \log \frac{D_{|\boldsymbol{v}|} + d_a}{D_{|\boldsymbol{v}|} + d_b},
\end{aligned}$$

*where (a) is because $N_a$ is independent of $S_i, i \in G_{\boldsymbol{v}}$; (b) is because conditioning reduces entropy; (c) is by applying the chain rule, and the facts that $N_a^n$ is an i.i.d. sequence and conditioning reduces entropy; (d) is by applying the mutual information game result (see page 263, [22], as well as [24]) that Gaussian noise is the worst additive noise under a variance constraint, and taking $N_a(i)$ as channel input; finally (e) is due to the convexity and monotonicity of $\log \frac{x+d_a}{x+d_b}$ in $x \in (0, \infty)$ when $d_a \geq d_b \geq 0$. This completes the proof for the second claim.*

We note that a similar line of argument was used in [8] to derive a sum rate lower bound for a system with two levels of distortion constraints. However, Lemma 3.1 generalizes that result since there exists only one auxiliary random variable in the setting of [8], but there are two auxiliary random variables $Y_a$ and $Y_b$ in the current setting.

# 9 Proof of Corollary 5.2

**Proof 6** *We first rewrite the rate formula given in Theorem 5.2. For a fixed set of (generalized symmetric) auxiliary random variables $\{\{Y_{\alpha,k}, \alpha \in I_{K-1}, k \in I_K\}, Y_K\}$, recall the definion the following quantities for*

*$\alpha \in I_{K-1}$*

$$\tilde{H}_\alpha(\boldsymbol{Y}) = h(Y_{\alpha,i}, i \in I_\alpha | Y_{j,k}, j \in I_{\alpha-1}, k \in I_\alpha) - \frac{\alpha}{K} h(Y_{\alpha,i}, i \in I_K | X, \{Y_{j,k}, j \in I_{\alpha-1}, k \in I_K\}), \quad (126)$$

*and*

$$\tilde{H}_K(\boldsymbol{Y}) = I(X; Y_K | Y_{\alpha,k}, \alpha \in I_{K-1}, k \in I_K). \quad (127)$$

*Then it follows that*

$$\sum_{\alpha=1}^{K} \frac{1}{\alpha} \tilde{H}_\alpha(\boldsymbol{Y}) = \sum_{\alpha=1}^{K-1} \frac{1}{\alpha} h(Y_{\alpha,i}, i \in I_\alpha | Y_{j,k}, j \in I_{\alpha-1}, k \in I_\alpha)$$
$$+ \frac{1}{K} h(Y_K | Y_{j,k}, j \in I_{K-1}, k \in I_K) - \frac{1}{K} h(\{Y_{j,k}, j \in I_{K-1}, k \in I_K\}, Y_K | X), \quad (128)$$

*where the right hand side is the rate expression given in Theorem 5.2.*

*Now for the specific set of random variables defined by (73) and (80), we have for $\alpha = 2, 3, ..., K-1$*

$$\begin{aligned}
\tilde{H}_\alpha(\boldsymbol{Y}) &= h(Y_{\alpha,i}, i \in I_\alpha | Y_{j,k}, j \in I_{\alpha-1}, k \in I_\alpha) \\
&\qquad - \frac{\alpha}{K} h(X + Z_{\alpha,i}, i \in I_K | X, \{X + Z_{j,k}, j \in I_{\alpha-1}, k \in I_K\}) \\
&\stackrel{(a)}{=} h(Y_{\alpha,i}, i \in I_\alpha | Y_{j,k}, j \in I_{\alpha-1}, k \in I_\alpha) - \frac{\alpha}{K} h(Z_{\alpha,i}, i \in I_K | Z_{j,k}, j \in I_{\alpha-1}, k \in I_K) \\
&\stackrel{(b)}{=} h(Y_{\alpha,i}, i \in I_\alpha | Y_{j,k}, j \in I_{\alpha-1}, k \in I_\alpha) - h(Z_{\alpha,i}, i \in I_\alpha | Z_{j,k}, j \in I_{\alpha-1}, k \in I_\alpha) \\
&\stackrel{(c)}{=} h(Y_{\alpha,i}, i \in I_\alpha | Y_{j,k}, j \in I_{\alpha-1}, k \in I_\alpha) - h(Y_{\alpha,i}, i \in I_\alpha | X, \{Y_{j,k}, j \in I_{\alpha-1}, k \in I_\alpha\}) \\
&= I(Y_{\alpha,i}, i \in I_\alpha; X | Y_{j,k}, j \in I_{\alpha-1}, k \in I_\alpha), \qquad (129)
\end{aligned}$$

*where (a) and (c) are because $X$ is independent of $Z_{\alpha,i}$; (b) is because of the chain rule and the fact that $Z_{\alpha,i}$ is independent of $\{Z_{\alpha,k}, \alpha \in I_K, k \neq i\}$. Because of the Markov string $\{Y_{1,k}, k \in I_K\} \leftrightarrow \{Y_{2,k}, k \in I_K\} \leftrightarrow ... \leftrightarrow \{Y_{K-1,k}, k \in I_K\} \leftrightarrow X$, we have*

$$\tilde{H}_\alpha(\boldsymbol{D}^*) = h(X | Y_{\alpha-1,i}, i \in I_\alpha) - h(X | Y_{\alpha,i}, i \in I_\alpha) = \frac{1}{2} \log \frac{(\alpha-1) D^*_{\alpha-1}}{(\alpha - D^*_{\alpha-1}) D^*_\alpha}, \quad (130)$$

*by the choices of the variances of the Gaussian random variables $N_{\alpha,k}$. For $\alpha = 1$ and $\alpha = K$, it is straightforward to verify that*

$$\begin{aligned}
\tilde{H}_1(\boldsymbol{D}^*) &= \frac{1}{2} \log \frac{1}{D^*_1}, \\
\tilde{H}_K(\boldsymbol{D}^*) &= \frac{1}{2} \log \frac{(K-1) D^*_{K-1}}{(K - D^*_{K-1}) D^*_K}. \qquad (131)
\end{aligned}$$

*Combining (130) and (131) we have,*

$$\begin{aligned}
\sum_{\alpha=1}^{K} \frac{1}{\alpha} \tilde{H}_\alpha(\boldsymbol{D}^*) &= \frac{1}{2} \log \frac{1}{D^*_1} + \frac{1}{2} \sum_{\alpha=2}^{K} \frac{1}{\alpha} \log \frac{(\alpha-1) D^*_{\alpha-1}}{(\alpha - D^*_{\alpha-1}) D^*_\alpha} \\
&= \frac{1}{2} \sum_{\alpha=1}^{K} \log \frac{D^*_{\alpha-1}}{D^*_\alpha} - \frac{1}{2} \sum_{\alpha=2}^{K} \frac{1}{\alpha} \log \frac{\alpha - D^*_{\alpha-1}}{\alpha - 1}, \qquad (132)
\end{aligned}$$

*which completes the proof by defining $D^*_0 \triangleq 1$.*

## 10 Proof of Corollary 5.3

**Proof 7** *To facilitate discussion, define the following index set of loose constraints*

$$C_L = \{\alpha : D^*_\alpha < D_\alpha\}, \tag{133}$$

*and it follows that $C_L^c = I_K \backslash C_L$; note that $1 \in C_L^c$. For a given $\alpha \in C_L$, define $N(\alpha)$ as the index of lower neighboring distortion constraint to $\alpha$ that is not loose, i.e., $N(\alpha) = \max_{k<\alpha, k\in C_L^c} k$.*

*We first consider the case when the distortion vector is given such that it satisfies the conditions*

$$\Phi_{\alpha-1}(D_{\alpha-1}) \geq \Phi_\alpha(D_\alpha), \quad \alpha = 2, 3, ..., K, \tag{134}$$

*where we take $D_0 \triangleq 1$. Note this implies $D_\alpha = D^*_\alpha$, $\alpha = 1, 2, ..., K$, and $C_L = \emptyset$. In this case we choose $d_\alpha = \Phi_\alpha(D_\alpha)$, for $\alpha = 1, 2, ..., K-1$, which is clearly valid. We start from Theorem 5.4 to show that for the specific choice of $d_\alpha$, the claims holds.*

$$\begin{aligned}
\sum_{i=1}^K R_i &\geq \frac{K}{2}\sum_{\alpha=1}^K \frac{1}{\alpha}\log\frac{(1+d_\alpha)(D_\alpha+d_{\alpha-1})}{(1+d_{\alpha-1})(D_\alpha+d_\alpha)}\\
&= \frac{K}{2}\sum_{\alpha=2}^{K-1}\frac{1}{\alpha}\log\left[\frac{D_{\alpha-1}}{D_\alpha}\frac{1+(\alpha-1)D_\alpha}{1+(\alpha-2)D_{\alpha-1}}\frac{\alpha-1-D_\alpha+\frac{D_\alpha}{D_{\alpha-1}}}{\alpha+1-D_\alpha}\right]\\
&\quad +\frac{K}{2}\log\frac{1}{D_1(2-D_1)}+\frac{1}{2}\log\left[\frac{D_{K-1}}{D_K}\frac{K-1-D_K+\frac{D_K}{D_{K-1}}}{1+(K-2)D_{K-1}}\right]\\
&= \frac{K}{2}\sum_{\alpha=1}^K\frac{1}{\alpha}\log\frac{D_{\alpha-1}}{D_\alpha}+\frac{K}{2}\sum_{\alpha=2}^{K-1}\frac{1}{\alpha}\log\left[\frac{1+(\alpha-1)D_\alpha}{1+(\alpha-2)D_{\alpha-1}}\frac{\alpha-1-D_\alpha+\frac{D_\alpha}{D_{\alpha-1}}}{\alpha+1-D_\alpha}\right]\\
&\quad +\frac{K}{2}\log\frac{1}{(2-D_1)}+\frac{1}{2}\log\left[\frac{K-1-D_K+\frac{D_K}{D_{K-1}}}{1+(K-2)D_{K-1}}\right]\\
&= \frac{K}{2}\sum_{\alpha=1}^K\frac{1}{\alpha}\log\frac{D_{\alpha-1}}{D_\alpha}+\frac{K}{2}\log\frac{1}{(2-D_1)}+\frac{K}{2}\sum_{\alpha=2}^{K-1}[\frac{1}{\alpha}-\frac{1}{\alpha+1}]\log\left(1+(\alpha-1)D_\alpha\right)\\
&\quad +\frac{K}{2}\sum_{\alpha=2}^{K-1}\frac{1}{\alpha}\log\frac{\alpha-1-D_\alpha+\frac{D_\alpha}{D_{\alpha-1}}}{\alpha+1-D_\alpha}+\frac{1}{2}\log\left(K-1-D_K+\frac{D_K}{D_{K-1}}\right)\\
&\overset{(a)}{\geq} \frac{K}{2}\sum_{\alpha=1}^K\frac{1}{\alpha}\log\frac{D_{\alpha-1}}{D_\alpha}+\frac{K}{2}\log\frac{1}{(2-D_1)}+\frac{K}{2}\sum_{\alpha=2}^{K-1}\frac{1}{\alpha}\log\frac{\alpha-1}{\alpha+1-D_\alpha}+\frac{1}{2}\log\left(K-1\right)\\
&= \frac{K}{2}\sum_{\alpha=1}^K\frac{1}{\alpha}\log\frac{D_{\alpha-1}}{D_\alpha}-\frac{K}{2}\sum_{\alpha=2}^K\frac{1}{\alpha-1}\log(\alpha-D_{\alpha-1})+\frac{K}{2}\sum_{\alpha=2}^K\frac{1}{\alpha}\log(\alpha-1)\\
&\geq \frac{K}{2}\sum_{\alpha=1}^K\frac{1}{\alpha}\log\frac{D_{\alpha-1}}{D_\alpha}-\frac{K}{2}\sum_{\alpha=2}^K\frac{1}{\alpha-1}\log\alpha+\frac{K}{2}\sum_{\alpha=2}^K\frac{1}{\alpha}\log(\alpha-1)
\end{aligned} \tag{135}$$

*where in (a) we used $\frac{D_\alpha}{D_{\alpha-1}} \geq D_\alpha$, and omitted the third term which is positive. Thus the claim is true if (134) holds.*

*For the case when (134) does not hold, then we choose $d_\alpha = \Phi_\alpha(D_\alpha)$, for $\alpha \in C_L^c$ as before; however for any $\alpha \in C_L$, we choose $d_\alpha = d_{N(\alpha)}$. Note that with such a choice, we have $d_\alpha = d_{\alpha-1}$ and therefore,*

$$\log \frac{(1+d_\alpha)(D_\alpha + d_{\alpha-1})}{(1+d_{\alpha-1})(D_\alpha + d_\alpha)} = 0, \quad \alpha \in C_L. \tag{136}$$

*If we replace $D_\alpha$ with $D_\alpha^*$ in the left hand side of the above equation, the equality still holds; moreover*

$$d_\alpha = \Phi_\alpha(D_\alpha^*), \quad \alpha \in C_L \tag{137}$$

*Thus by using this particular choice of $(d_1, d_2, ..., d_{K-1})$, we have*

$$\begin{aligned} \sum_{i=1}^{K} R_i &\geq \frac{K}{2} \sum_{\alpha=1}^{K} \frac{1}{\alpha} \log \frac{(1+d_\alpha)(D_\alpha + d_{\alpha-1})}{(1+d_{\alpha-1})(D_\alpha + d_\alpha)} \\ &= \frac{K}{2} \sum_{\alpha=1}^{K} \frac{1}{\alpha} \log \frac{(1+d_\alpha)(D_\alpha^* + d_{\alpha-1})}{(1+d_{\alpha-1})(D_\alpha^* + d_\alpha)}, \end{aligned} \tag{138}$$

*and the exact same derivation holds as in the case when (134) holds, with $D_\alpha^*$ replacing $D_\alpha$. Dividing both side of (135) by $K$ completes the proof of the corollary.*

# 11 Proof of Theorem 5.5

**Proof 8** *We pick up the story from (83) for the lower bound and rewrite it slightly differently.*

$$\begin{aligned} n\sum_{i=1}^{K}(R_i+\epsilon) \geq \sum_{\alpha=1}^{K-1} \frac{K}{\alpha\binom{K}{\alpha}} \sum_{G_{\boldsymbol{v}}:|\boldsymbol{v}|=\alpha} &\left[I(S_i, i \in G_{\boldsymbol{v}}; Y_\alpha^n) - I(S_i, i \in G_{\boldsymbol{v}}; Y_{\alpha-1}^n)\right] \\ &+ \left[I(S_i, i \in I_K; X^n) - I(S_i, i \in I_K; Y_{K-1}^n)\right]. \end{aligned} \tag{139}$$

*where now the random variables $Y_\alpha$, $\alpha = 1, 2, ..., K-1$ are defined as in (66) and (67), and for simplicity we define $Y_0 = 0$, i.e., a constant.*

*Next we consider the upper bound $\hat{R}'(\boldsymbol{D})$, c.f. Theorem 6.1, using the same set of random variables $Y_\alpha$, $\alpha = 1, 2, ..., K$ as above*

$$\hat{R}'(\boldsymbol{D}) = \sum_{\alpha=1}^{K} \frac{1}{\alpha} I(X; Y_\alpha | Y_{\alpha-1}) = \sum_{\alpha=1}^{K} \frac{1}{\alpha} [I(X; Y_\alpha) - I(X; Y_{\alpha-1})]. \tag{140}$$

*Note we have used that fact that $X \leftrightarrow Y_\alpha \leftrightarrow Y_{\alpha-1}$ is a Markov string for any $\alpha \in I_K$. The auxiliary random variables used in the lower and upper bounds are in fact the same, and it is clear that this is a valid choice in deriving the lower bound by definition.*

*Thus we can now bound the difference between the upper and lower bound on the symmetric individual-description rate as follows*

$$
\begin{aligned}
&\sum_{\alpha=1}^{K}\frac{1}{\alpha}I(X;Y_\alpha|Y_{\alpha-1})-\frac{1}{nK}\left[I(S_i,i\in I_K;X^n)-I(S_i,i\in I_K;Y^n_{K-1})\right]\\
&\qquad -\sum_{\alpha=1}^{K-1}\left[\frac{1}{n\alpha\binom{K}{\alpha}}\sum_{G_{\boldsymbol{v}}:|\boldsymbol{v}|=\alpha}\left[I(S_i,i\in G_{\boldsymbol{v}};Y^n_\alpha)-I(S_i,i\in G_{\boldsymbol{v}};Y^n_{\alpha-1})\right]\right]\\
=&\sum_{\alpha=1}^{K-1}\left\{\frac{1}{\alpha\binom{K}{\alpha}}\sum_{G_{\boldsymbol{v}}:|\boldsymbol{v}|=\alpha}\left[I(X;Y_\alpha)-I(X;Y_{\alpha-1})-\frac{1}{n}I(S_i,i\in G_{\boldsymbol{v}};Y^n_\alpha)+\frac{1}{n}I(S_i,i\in G_{\boldsymbol{v}};Y^n_{\alpha-1})\right]\right\}\\
&+\frac{1}{K}\left\{I(X;Y_K)-I(X;Y_{K-1})-\frac{1}{n}I(S_i,i\in I_K;X^n)+\frac{1}{n}I(S_i,i\in I_K;Y^n_{K-1})\right\} \qquad (141)
\end{aligned}
$$

*Now consider an arbitrary $\alpha\in I_{K-1}$, and an arbitrary $\boldsymbol{v}$ such that $|\boldsymbol{v}|=\alpha$, it follows that*

$$
\begin{aligned}
&I(X;Y_\alpha)-I(X;Y_{\alpha-1})-\frac{1}{n}I(S_i,i\in G_{\boldsymbol{v}};Y^n_\alpha)+\frac{1}{n}I(S_i,i\in G_{\boldsymbol{v}};Y^n_{\alpha-1})\\
=\;&h(Y_\alpha)-h(Y_\alpha|X)-h(Y_{\alpha-1})+h(Y_{\alpha-1}|X)\\
&\qquad -\frac{1}{n}[h(Y^n_\alpha)-h(Y^n_\alpha|S_i,i\in G_{\boldsymbol{v}})]+\frac{1}{n}[h(Y^n_{\alpha-1})-h(Y^n_{\alpha-1}|S_i,i\in G_{\boldsymbol{v}})]\\
\stackrel{(a)}{=}\;&-h(Z_\alpha)+h(Z_{\alpha-1})+\frac{1}{n}h(Y^n_\alpha|S_i,i\in G_{\boldsymbol{v}})-\frac{1}{n}h(Y^n_{\alpha-1}|S_i,i\in G_{\boldsymbol{v}})\\
\stackrel{(b)}{=}\;&\frac{1}{2}\log\frac{d_{\alpha-1}}{d_\alpha}-\frac{1}{n}I(Y^n_{\alpha-1};N^n_{\alpha-1}|S_i,i\in G_{\boldsymbol{v}})\\
\stackrel{(c)}{\leq}\;&\frac{1}{2}\log\left[\frac{d_{\alpha-1}}{d_\alpha}\frac{D_\alpha+d_\alpha}{D_\alpha+d_{\alpha-1}}\right]\\
=\;&\frac{1}{2}\log\frac{2-D_\alpha}{1-D_\alpha+\frac{D_\alpha}{D_{\alpha-1}}}\stackrel{(c)}{\leq}\frac{1}{2} \qquad (142)
\end{aligned}
$$

*where (a) is due to $Y^n_\alpha$ and $Y^n_{\alpha-1}$ are independent squences, (b) holds since $Y_{\alpha-1}=Y_\alpha+N_{\alpha-1}$, and in (c) we used the bounding technique used in the proof of Lemma 3.1, and continued to use the definition of*

$$
d_\alpha=\sum_{i=\alpha}^{K}\sigma_i^2=\frac{D_\alpha}{1-D_\alpha},\quad \alpha=1,2,...,K-1, \qquad (143)
$$

*and finally in (c) we used the fact $D_\alpha-\frac{D_\alpha}{D_{\alpha-1}}\leq 0$ and $D_\alpha\geq 0$.*

*The last term in (141) can be bounded similarly by noticing $I(S_i,i\in I_K;X^n)\geq I(S_i,i\in I_K;Y^n_K)$*

$$
\begin{aligned}
&I(X;Y_K)-I(X;Y_{K-1})-\frac{1}{n}I(S_i,i\in I_K;X^n)+\frac{1}{n}I(S_i,i\in I_K;Y^n_{K-1})\\
\leq\;&I(X;Y_K)-I(X;Y_{K-1})-\frac{1}{n}I(S_i,i\in I_K;Y^n_K)+\frac{1}{n}I(S_i,i\in I_K;Y^n_{K-1})\\
=\;&-h(Z_K)+h(Z_{K-1})+\frac{1}{n}h(Y^n_K|S_i,i\in I_K)-\frac{1}{n}h(Y^n_{K-1}|S_i,i\in I_K)\\
\leq\;&\frac{1}{2}\log\left[\frac{d_{K-1}}{\sigma_K^2}\frac{D_K+\sigma_K^2}{D_K+d_{K-1}}\right]\leq\frac{1}{2} \qquad (144)
\end{aligned}
$$

*where the last step is by $\sigma_K^2 = \frac{D_K}{1-D_K}$.*

*Now summarize all the bounds derived above, we have that*

$$\hat{R}(\boldsymbol{D}) - R(\boldsymbol{D}) \leq \sum_{\alpha=1}^{K} \frac{1}{2\alpha}, \tag{145}$$

*which completes the proof.*

# 12 Proof of Theorem 6.2

**Proof 9 (Proof of Theorem 6.2)** *Fix a set of (generalized symmetric) random variables $\{\{Y_{\alpha,k}, \alpha \in I_{K-1}, k \in I_K\}, Y_K\}$. For a given $\boldsymbol{A} \geq 0$, let $l_\alpha$ be the non-negative integer defined in Lemma 2.2 for the $\alpha$-level. For any $\alpha \in I_K$, let*

$$R_{\alpha,k} = \begin{cases} 0 & \text{if} \quad 1 \leq k \leq l_\alpha; \\ \frac{\tilde{H}_\alpha}{\alpha - l_\alpha} & \text{if} \quad l_\alpha + 1 \leq k \leq K. \end{cases} \tag{146}$$

*$R_{\alpha,k}$ will be the rate assigned to the $\alpha$-th layer for the $k$-th description; denote $(R_{\alpha,1}, R_{\alpha,2}, ..., R_{\alpha,K})$ as $\boldsymbol{R_\alpha}$. It is clear from the original PPR multilayer scheme [6] that if each of the description has rate approximately $\tilde{H}_\alpha/\alpha$ at the $\alpha$-th level, then any of the $\alpha$ descriptions can guarantees decoding with high probability. However, because the first $l_\alpha$ descriptions are not given any rate for the $\alpha$-th layer in (146), this can not be achieved directly without proper coding.*

*The generalized coding scheme is by combining the original PPR multlayer scheme with proper MDS channel codes. The PPR multilayer scheme is still used as the main encoding step, and let us denote the codeword (the output index written in a large enough appropriate alphabet) for the $\alpha$-th level for description $k$ as $C_{\alpha,k}$, for $\alpha \in I_K$. A post-coding packaging step is now added at the $\alpha$-th layer as follows. The last $K - l_\alpha$ codeword indices are written in the descriptions as in the original scheme. Each of the first $l_\alpha$ codeword indice $C_{\alpha,k}$, $k = 1, 2, ..., l_\alpha$ is encoded by a ($K - l_\alpha$, $\alpha - l_\alpha$) MDS code, and each of the resulting codeword (index) is written into one of the last $K - l_\alpha$ description. This results in an additional rate $\tilde{H}_\alpha/\alpha(\alpha - l_\alpha)$ in each description. Note that since $l_\alpha \leq \alpha - 1$, the above MDS code rate is always well defined. It is clear that the rate of the $k$-th description, $k > l_\alpha$, for the $\alpha$-th layer is*

$$R_{\alpha,k} = \frac{\tilde{H}_\alpha}{\alpha} + \frac{\tilde{H}_\alpha}{\alpha(\alpha - l_\alpha)} * l_\alpha = \frac{\tilde{H}_\alpha}{\alpha - l_\alpha}, \tag{147}$$

*as we claimed.*

*At the decoder, suppose $k$ descriptions in the set $G_{\boldsymbol{v}}$ are available, where $|\boldsymbol{v}| = k$. Consider a specific level $\alpha \in I_k$, and the pre-decoding unpackaging procedure is as follows. Suppose $n_\alpha$ of indices in $G_{\boldsymbol{v}}$ is smaller or equal to $l_\alpha$, i.e., $n_\alpha = |G_{\boldsymbol{v}} \cap I_{l_\alpha}|$. In the remaining $k - n_\alpha$ descriptions, clearly we can recover their respective codewords, i.e., $C_{\alpha,i}$ for $i \in G_{\boldsymbol{v}} \setminus I_{l_\alpha}$. However, since $n_\alpha \leq l_\alpha$, we have also $k - n_\alpha \geq \alpha - n_\alpha \geq \alpha - l_\alpha$ pieces of the MDS encoded $C_{\alpha,i}$ for $i \in I_{l_\alpha}$, which can be correctly decoded by the property of the MDS code. Since $G_{\boldsymbol{v}} \cap I_{l_\alpha} \subseteq I_{l_\alpha}$, we can recover all $C_{\alpha,i}$ for $i \in G_{\boldsymbol{v}}$. This holds true for all $\alpha = 1, 2, ..., k$, and then the main decoding step in the PPR multilayer scheme can be applied.*

*We remark here that the decoding can be easily improved, because if $n_\alpha < l_\alpha$, there is additional information that the main decoding step is not utilizing. However the above simple procedure suffices for proving the current theorem.*

*It remains to show (113) is true with the given rate vector, the proof of which follows closely the step in [16] for the proof of Theorem 6.3. Let $\{c(\boldsymbol{v})\}$ be an optimal $\alpha$-resolution for $\boldsymbol{A}$. We have*

$$\boldsymbol{A} \cdot \boldsymbol{R_\alpha} = \sum_{\boldsymbol{v} \in \Omega_K^\alpha} c_\alpha(\boldsymbol{v})(\boldsymbol{v} \cdot \boldsymbol{R_\alpha}) + (\boldsymbol{A} - \sum_{\boldsymbol{v} \in \Omega_K^\alpha} c_\alpha(\boldsymbol{v})\boldsymbol{v}) \cdot \boldsymbol{R_\alpha}.$$

*By Lemma 2.1, for any $\boldsymbol{v}$ where $|\boldsymbol{v}| = \alpha$ such that $c_\alpha(\boldsymbol{v}) > 0$, $v_i = 1$ for $i = 1, 2, ..., l_\alpha$; moreover, exactly $\alpha - l_\alpha$ of the remaining components are $1$'s. Since the first $l_\alpha$ components of $\boldsymbol{R_\alpha}$ are $0$'s, and the remaining components are equal, we have*

$$\boldsymbol{v} \cdot \boldsymbol{R_\alpha} = (\alpha - l_\alpha)\frac{\tilde{H}_\alpha}{\alpha - l_\alpha} = \tilde{H}_\alpha \qquad \textit{for} \qquad \boldsymbol{v} : c_\alpha(\boldsymbol{v}) > 0.$$

*It follows that*

$$\sum_{\boldsymbol{v} \in \Omega_K^\alpha} c_\alpha(\boldsymbol{v})(\boldsymbol{v} \cdot \boldsymbol{R_\alpha}) = \tilde{H}_\alpha \sum_{\boldsymbol{v} \in \Omega_K^\alpha} c_\alpha(\boldsymbol{v}) = f_\alpha(\boldsymbol{A})\tilde{H}_\alpha. \tag{148}$$

*Since*

$$\boldsymbol{A} - \sum_{\boldsymbol{v} \in \Omega_K^\alpha} c_\alpha(\boldsymbol{v})\boldsymbol{v} = \boldsymbol{A} - \breve{\boldsymbol{A}} \tag{149}$$

*has zeros in the last $K - l_\alpha$ components, and $\boldsymbol{R_\alpha}$ has zeros in the complement positions, we have*

$$(A - \sum_{\boldsymbol{v} \in \Omega_K^\alpha} c_\alpha(\boldsymbol{v})\boldsymbol{v}) \cdot \boldsymbol{R_\alpha} = 0. \tag{150}$$

*It follows*

$$\boldsymbol{A} \cdot \boldsymbol{R_\alpha} = f_\alpha(\boldsymbol{A})\tilde{H}_\alpha. \tag{151}$$

*Summing over $\alpha \in I_K$ now completes the proof since $R_i = \sum_{\alpha=1}^{K} R_{\alpha,i}$.*

## 13 Proof of Corollary 6.2

**Proof 10** *Follow the proof approach for Corollary 5.3, however we directly use* $\boldsymbol{D^*}$ *to replace* $\boldsymbol{D}$*. Let* $d_\alpha = \Phi_\alpha(D^*_\alpha)$ *for* $\alpha = 1, 2, ..., K-1$*, we have*

$$\begin{aligned}
\sum_{i=1}^{K} A_i R_i &\geq \frac{1}{2}\sum_{\alpha=2}^{K-1} f_\alpha(\boldsymbol{A})\left[\log\frac{D^*_{\alpha-1}}{D^*_\alpha} + \log\frac{1+(\alpha-1)D^*_\alpha}{1+(\alpha-2)D^*_{\alpha-1}} + \log\frac{\alpha-1-D^*_\alpha+\frac{D^*_\alpha}{D^*_{\alpha-1}}}{1+\alpha-D^*_\alpha}\right] \\
&\quad + \frac{1}{2}f_1(\boldsymbol{A})\log\frac{1}{D^*_1(2-D^*_1)} + \frac{1}{2}f_K(\boldsymbol{A})\log\left[\frac{D^*_{K-1}}{D^*_K}\frac{K-1-D^*_K+\frac{D^*_K}{D^*_{K-1}}}{1+(K-2)D^*_{K-1}}\right] \\
&\overset{(a)}{\geq} \frac{1}{2}\sum_{\alpha=1}^{K} f_\alpha(\boldsymbol{A})\log\frac{D^*_{\alpha-1}}{D^*_\alpha} + \frac{1}{2}\sum_{\alpha=2}^{K-1}[f_{\alpha-1}(\boldsymbol{A}) - f_\alpha(\boldsymbol{A})]\log(1+(\alpha-1)D^*_\alpha) \\
&\quad - \frac{1}{2}\sum_{\alpha=2}^{K} f_{\alpha-1}(\boldsymbol{A})\log(\alpha-D^*_{\alpha-1}) + \frac{1}{2}\sum_{\alpha=2}^{K} f_\alpha(\boldsymbol{A})\log(\alpha-1) \\
&\overset{(b)}{\geq} \frac{1}{2}\sum_{\alpha=1}^{K} f_\alpha(\boldsymbol{A})\log\frac{D^*_{\alpha-1}}{D^*_\alpha} - \frac{1}{2}\sum_{\alpha=2}^{K} f_{\alpha-1}(\boldsymbol{A})\log(\alpha-D^*_{\alpha-1}) + \frac{1}{2}\sum_{\alpha=2}^{K} f_\alpha(\boldsymbol{A})\log(\alpha-1), \qquad (152)
\end{aligned}$$

*where (a) is true because* $\frac{D_\alpha}{D_{\alpha-1}} \geq D_\alpha$*, and in (b) we omitted the second term, because Lemma 2.3 implies* $f_\alpha(\boldsymbol{A}) \leq \frac{\alpha-1}{\alpha}f_{\alpha-1}(\boldsymbol{A}) \leq f_{\alpha-1}(\boldsymbol{A})$*. This completes the proof.*

## 14 Proof of Corollary 6.3

**Proof 11** *Clearly we only need to prove that the set* $\mathcal{R}_L(\boldsymbol{D}^*)$ *is a polytope. Since* $\log(\alpha - D^*_{\alpha-1}) \geq 0$ *for* $\alpha \geq 2$*, we can construct a set of independent fictitious source* $U_1, U_2, ..., U_K$*, such that*

$$H(U_\alpha) = \frac{1}{2}\log(\alpha+1-D^*_\alpha), \quad \alpha = 1, 2, ..., K-1, \qquad (153)$$

*and* $H(U_K) = 0$*. The MLD coding rate region for this* $K$*-source can be equivalently given in two forms, as implied by Theorem 6.3, with* $\tilde{H}_\alpha(\boldsymbol{Y})$ *replaced by* $H(U_\alpha)$*. Since the rate region of this MLD coding problem is clearly a polytope, there exists a finite set of rate vectors, denoted as* $\mathcal{S}_r$*, such that for any* $\boldsymbol{A}$*, there exists at least one rate vector* $(r_1, r_2, ..., r_K) \in \mathcal{S}_r$*, such that*

$$\sum_{i=1}^{K} A_i r_i = \frac{1}{2}\sum_{\alpha=2}^{K} f_{\alpha-1}(\boldsymbol{A})\log(\alpha - D^*_{\alpha-1}) \qquad (154)$$

*Now define* $\grave{R}_i = R_i + r_i$*,* $i = 1, 2, ..., K$*, and consequently (118) reduces to the condition that*

$$\sum_{i=1}^{K} A_i \grave{R}_i \geq \frac{1}{2}\sum_{\alpha=1}^{K} f_\alpha(\boldsymbol{A})\log\frac{D^*_{\alpha-1}}{D^*_\alpha} + \frac{1}{2}\sum_{\alpha=2}^{K} f_\alpha(\boldsymbol{A})\log(\alpha-1), \qquad (155)$$

*We can again define a set of fictitious independent sources $W_1, W_2, ..., W_K$, such that*

$$H(W_\alpha) = \log \frac{D^*_{\alpha-1}}{D^*_\alpha} + \log(\alpha - 1), \quad \alpha = 2, 3, ..., K, \tag{156}$$

*and*

$$H(W_1) = \log \frac{1}{D^*_\alpha}. \tag{157}$$

*Now we would like to apply Theorem 6.3 to assert (155) is in fact a characterization of the MLD coding rate region for this source, however one technicality has to be addressed first. Recall that $\boldsymbol{R}$ is not constrained to be non-negative, because otherwise $\grave{\boldsymbol{R}}$ must satisfy the additional constraint $\grave{\boldsymbol{R}} \geq \boldsymbol{r}$, and Theorem 6.3 can not be applied directly. However, by relaxing $\boldsymbol{R}$ to allow negative component, $\grave{\boldsymbol{R}}$ may have non-positive components, which will render Theorem 6.3 not applicable without the fact given in the remark immediately after Theorem 6.3. With that remark, now by applying Theorem 6.3, we see that (155) is indeed a characterization of the MLD coding rate region for this source.*

*Since the MLD coding rate region is a polytope, there exists a finite set of rate vectors $\mathcal{S}_{\grave{R}}$ such that for any $\boldsymbol{A}$, there exists at least one rate vector $(\grave{R}_1, \grave{R}_2, ..., \grave{R}_1) \in \mathcal{S}_{\grave{R}}$, such that (155) is satisfied with equality. Since both $\mathcal{S}_r$ and $\mathcal{S}_{\grave{R}}$ are finite, it follows that there exists a finite set $S_R$, such that for any $\boldsymbol{A}$, there exists at least one vector $\boldsymbol{R} = \grave{\boldsymbol{R}} - \boldsymbol{r} \in \mathcal{S}_R$ satisfying (118) with equality. This subsequently implies that the set $\mathcal{R}_L(\boldsymbol{D}^*)$ is a polytope, which completes the proof.*